# Ultrafast Liquid Jet X-ray Total Scattering

Adam F. Sapnik,[a*] Philip A. Chater,[b] Dean S. Keeble,[b] Tobias A. Bird,[b] John S. O. Evans,[c] Federica Bertolotti,[d] Antonietta Guagliardi,[e] Norberto Masciocchi,[d] Daniel T. Bowron,[f] Lise J. Støckler,[g] Elodie A. Harbourne,[h] Anders B. Borup,[g] Rebecca S. Silberg,[a] Viktoria A. Kaznelson,[g] Stefanie Punke,[a] Mariia Svyrydenko,[i,j] Maryna I. Bodnarchuk,[i,j] Dmitry N. Dirin,[i,j] Simon C. Boehme,[j,i] Maksym V. Kovalenko,[i,j] Bo B. Iversen,[g] Emil S. Bozin,[k] Kirsten M. Ø. Jensen,[a] Gabriele Ansaldi,[l] Karen Appel,[l] Alexander Gierke,[l,m] Jörg Hallmann,[l] Aliaksandr Leonau,[l] Anders Madsen,[l] Johannes Möller,[l] Jan-Etienne Pudell,[l] Angel Rodriguez-Fernandez,[l] Andreas Schmidt,[l] Roman Shayduk,[l] Cornelius Strohm,[m] Joana Valerio,[l] Ulrike Boesenberg,[l] and David A. Keen.[f*]

Email: afs@chem.ku.dk & david.keen@stfc.ac.uk

---

## Abstract

X-ray free-electron lasers (XFELs) provide the temporal resolution needed to extend total scattering and pair distribution function (PDF) analysis to investigate structural dynamics on femtosecond-to-picosecond timescales. We recently demonstrated that high-quality total scattering data and PDFs can be obtained from individual XFEL pulses; here, we take the next step towards time-resolved studies of structural dynamics in solution. Using the Materials Imaging and Dynamics (MID) instrument at the European XFEL, we obtain total scattering data up to $Q_{\mathrm{max}} = 16.4$ Å$^{-1}$ from both conventional capillary samples and, critically, from samples delivered by a liquid jet. Despite the large air and solvent backgrounds inherent to this setup, PDFs suitable for quantitative structural analysis are recovered from liquid dispersions and molecular clusters in solution. We further demonstrate pump–probe total scattering from $CsPbBr_3$ nanocrystals in a liquid jet, resolving a fluence-dependent photoinduced isotropic lattice strain. Additionally, empirical potential structure refinement (EPSR) of XFEL total scattering data from an aqueous Keggin cluster reveals site-specific hydration environments and enables three-dimensional mapping of the surrounding solvent structure. Together, these advances open the door to tracking not only ultrafast structural changes in solution, but also the evolving solvent environments that govern molecular behaviour and reactivity.

## Introduction

Total scattering and pair distribution function (PDF) analysis are powerful methods for determining atomic-scale structure in materials where local order is essential to function.[1] By including both Bragg and diffuse scattering, total scattering provides access to structural correlations beyond the average crystallographic unit cell, making it applicable to crystalline, nanocrystalline, amorphous, liquid, and molecular or cluster-containing systems.[2] The PDF, obtained by Fourier transforming appropriately normalised total-scattering data, provides a real-space representation of interatomic distances, from which local bonding, coordination environments, disorder, and structural coherence lengths can be extracted.[3] However, meaningful PDF analysis requires sufficiently high-quality data collected to high momentum transfer, $Q$, because the maximum measured $Q$, $Q_{\max}$, directly determines the real-space resolution ($\Delta r \approx \pi/Q_{\max}$) and because noise at high $Q$ is amplified during PDF generation.[4] High-quality, high-$Q$ data are therefore essential for quantitative structural analysis.

Synchrotron X-ray sources remain the gold standard for X-ray total scattering and PDF measurements. Dedicated high-energy beamlines routinely deliver data to high $Q_{\max}$, enabling high-resolution PDFs suitable for detailed modelling of complex local structure.[1] The principal limitation of synchrotron measurements is their temporal resolution. Although modern beamlines can collect high-quality total scattering data on timescales of seconds, and in some cases below one second, many chemical processes occur far more rapidly.[5] Photoinduced reactions, non-thermal structural rearrangements, ultrafast phase transitions, and transient molecular or nanoscale intermediates often evolve on femtosecond-to-picosecond timescales, beyond the reach of conventional synchrotron PDF experiments.[6–10]

X-ray free-electron lasers (XFELs) provide the femtosecond X-ray pulses required to access this ultrafast regime.[11] In principle, they offer an ideal route to time-resolved total scattering and PDF analysis, combining atomic-scale structural sensitivity with femtosecond temporal resolution. In practice, however, experimental geometry, detector coverage, background control, and the relatively modest photon energies available at many instruments have constrained XFEL total scattering.[6,10,12–15] These factors have typically limited either the accessible $Q_{\max}$, the quality of the normalised scattering data, or both. Despite this, XFEL scattering and PDF methods have been successfully applied to selected ultrafast structural problems, including photoinduced Au–Au bond formation in molecular clusters,[6] the suppression of Ir–Ir dimerisation during an ultrafast phase transition,[10] and the characterisation of liquids under extreme pressure and temperature.[13,14] In these cases, we can resolve pronounced structural rearrangements or broad scattering features despite the restricted *Q*-range; however, the resulting real-space resolution remains insufficient to routinely distinguish closely spaced atomic correlations and subtle local distortions, as is possible from synchrotron measurements.

In our recent work, we demonstrated that this limitation can be overcome using the High Energy Density (HED) instrument at the European XFEL.[16] Using a purpose-designed detector geometry and data-reduction workflow, we obtained high-quality, normalised, total-scattering data for $0.35 < Q < 16.6$ Å$^{-1}$ and generated robust PDFs from a wide range of materials, including crystalline powders, nanocrystalline samples, amorphous solids, liquids and clusters in solution. Crucially, we showed that a single ~30 fs XFEL pulse can yield quantitatively useful PDFs, establishing that XFELs can deliver near-synchrotron-quality local-structure data on femtosecond timescales. Building on this proof of principle, we now broaden the scope and versatility of ultrafast total scattering by extending the technique to more complex experimental environments and sample-delivery systems. The Materials Imaging and Dynamics (MID) instrument at the European XFEL provides the flexibility to achieve this, enabling liquid-jet measurements and time-resolved optical pump–probe experiments. Whereas our previous work at HED focused on establishing high-quality total scattering from predominantly static samples, the present study advances the technique toward probing structural dynamics in liquids and other condensed matter under experimentally relevant conditions. We benchmark high-*Q* total-scattering and PDF measurements at MID across capillary and liquid-jet sample geometries using crystalline standards, an amorphous solid, pure solvent scattering, a liquid dispersion and a molecular cluster in solution. Moreover, we extend beyond static measurements to include time-resolved pump–probe total scattering and experimentally constrained atomistic modelling to elucidate solute–solvent interactions in real space. In doing so, we demonstrate MID's capability for high-quality PDF analysis and assess remaining experimental challenges, particularly background subtraction, air scattering, and liquid-jet reproducibility, which must be properly addressed to enable time-resolved real-space studies of structural dynamics.

# Methods

**Instrument:** Our experiments were performed using the MID instrument at the European XFEL in Hamburg.[17,18] General details about MID are available in Ref. 19. We used a Varex 4343CT detector positioned to maximise the $2\theta$-range and provide a $Q$-range of 0.95–16.4 Å$^{-1}$ at 24.37 keV, with a beam energy of ~150 µJ and repetition rate of 10 Hz (**Fig. 1a & b**). The $Q_{max}$ yields an approximate real-space resolution of 0.19 Å ($\pi/Q_{max}$), nearly identical to our previous HED experiments.[16]

The 24 keV SASE beam was collimated by a first set of beryllium Compound Refractive Lenses (CRLs) and transported by two horizontal offset mirrors using the Pt stripe to the experimental station. A feedback loop using a 40 µm diamond imager at the end of the tunnel corrected for slow drifts. Focusing to a beam size of approximately 60 µm was achieved by a second set of beryllium CRLs located approximately 28 m upstream of the sample. A 200 µm single-crystalline diamond window upstream of the sample separated the beamline vacuum from the in-air experiment setup.

**Detector:** The Varex 4343CT detector has a 2880 × 2880 array of 150-µm-square pixels and was operated at 10 Hz. It was placed at 14.7° from the horizontal plane, with the point of nearest incidence at 138.93 mm, covering scattering angles between 4.5 and 85° in the vertical plane. Further details on the benefits of this tilted detector setup are provided in Refs. 20 and 16.

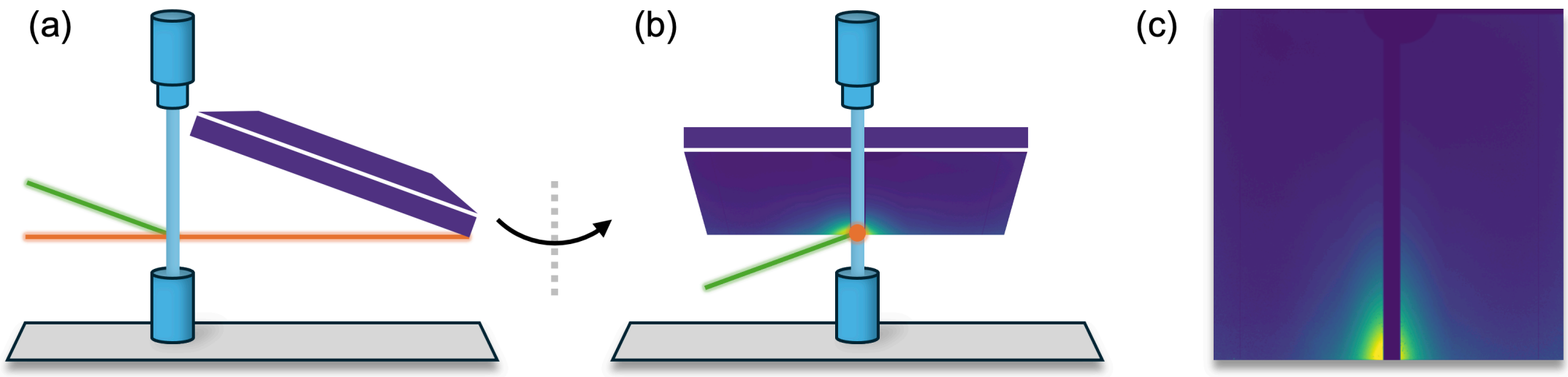

**Figure 1** Schematic of the experimental setup viewed **(a)** perpendicular to and **(b)** along the path of the XFEL beam: XFEL beam (orange), optical laser (green), liquid jet (light blue), and tilted Varex detector (dark blue). The XFEL beam and optical laser are arranged horizontally and coplanar, intersecting at the liquid jet. Other components of the setup have been omitted for clarity. **(c)** An example corrected detector image before integration.

**Sample Delivery:** Powder samples were loaded into fused silica capillaries (0.3 mm outer diameter) and mounted vertically. For liquid samples, a Shimadzu HPLC pump (LC-20AD xr) connected to a 5 mL metal reservoir delivered the sample through the liquid jet (**Fig. 1a & b**). Two modes of operation were used: single pass or recycling. In single-pass mode, the spent sample was collected in a single-use catcher before disposal, so each measurement used fresh sample. In recycling mode, the sample circulated through the liquid jet before being repumped through the HPLC system, which was used when the same sample could be measured repeatedly. To protect the detector and optics from stray liquids, the jet passed through a cylindrical Kapton shield with small holes to allow X-ray and laser transmission.

The liquid jet was generated using a 3D-printed gas dynamic virtual nozzle (GDVN; E-type) developed at EuXFEL.[21,22] These nozzles are highly versatile, with a sample orifice inner diameter of 180 μm, a gas orifice diameter of 145 μm, and an 180 μm separation between the orifices. This geometry enables tunable jet diameters of 10–100 μm by adjusting the helium gas pressure and sample flow rate (**Supplementary Table 1**). For single-pass operation, jet diameters of approximately 80–100 μm were achieved using a sample flow rate of 1 mL $min^{-1}$ and a helium gas pressure of 1379 kPa. For samples operated in recycling mode, jet diameters of approximately 75–100 μm were achieved using sample flow rates of 700–1200 μL $min^{-1}$ and helium gas flow rates of 15–18 mg $min^{-1}$.

**Optical Laser:** The pump–probe (PP) laser provides 50 fs 800 nm pulses that are synchronised to the X-ray delivery.[23] The laser was frequency-doubled to 400 nm and focused to 0.18 × 0.26 $mm^2$ to excite the sample. The pulse energy was controlled by a waveplate-polariser combination.

We performed pump–probe measurements on $CsPbBr_3$ nanocrystals at delay times of + and −5 ps using pump fluences of 0.82, 1.97, 3.94, and 5.77 mJ $cm^{-2}$. Laser excitation parameters were selected on the basis of previous ultrafast electron diffraction experiments performed on $FAPbBr_3$ nanocrystals of comparable size and adjusted during the experiment to manage the optical density (OD) of the suspension (OD ~ 1.5 for a jet thickness of 80 μm being an acceptable compromise).[24] The pump wavelength of 400 nm was selected to ensure above-bandgap photoexcitation.

**Data Processing**

We reduced the raw detector images to one-dimensional scattering patterns using a custom Python workflow. The reduction strategy involved detector calibration, image correction and normalisation, averaging, azimuthal integration, and subsequent data analysis.[16] The detector calibration was performed using pyFAI, and a fused silica capillary loaded with NIST $CeO_2$ SRM 674b powder was used as the calibrant.[25] We allowed the primary beam energy and sample-to-detector distance to refine simultaneously. We calculated a mean dark image from a separate dark run acquired without X-ray exposure and subtracted it from each sample image. A combined detector mask was constructed from a pre-existing mask, additional manually excluded detector regions (including those directly behind the liquid jet), the manufacturer bad-pixel mask, and pixels with invalid gain or solid-angle correction values (**Supplementary Fig. 1**). A gain correction to account for pixel-to-pixel variations in detector response and a solid-angle correction were then applied. Incident XFEL intensity fluctuations were corrected using the upstream X-ray gas monitor (XGM) signal. For each X-ray pulse train, the XGM value corresponding to the recorded X-ray pulse was extracted and compared with the run median. Trains with missing, non-finite or very weak (below 20% of the median) XGM signals were rejected. For accepted trains, detector images were normalised by the relative incident intensity (**Fig. 1c**). Finally, the corrected images were averaged and azimuthally integrated using pyFAI to obtain the one-dimensional scattering pattern.

**Data Analysis**

Rietveld refinement was performed using TOPAS Academic v8.[26] Reciprocal space analysis was performed using the Debye Scattering Equation (DSE) approach with the DebUsSy program suite.[27] Data from $CsPbBr_3$ nanocrystals suspended in toluene were modelled in the $Q$-range of 1.4–7.4 Å$^{-1}$. We used the solvent scattering signal, collected under the same experimental conditions as the sample, as the blank trace for DSE model refinement without prior subtraction of air scattering, and scaled it to the measured data by assuming air scattering contributes at the same scale. Following the DebUsSy strategy, atomistic models of $CsPbBr_3$ cuboidal nanocrystals with increasing sizes along two orthogonal growth directions were built to calculate the Gaussian-sampled interatomic distance driving the DSE.[28] The refined parameters included the nanocrystal and blank-trace scale factors, an isotropic lattice strain parameter, and isotropic Debye-Waller factors. PDFs were generated from the total scattering data using PDFgetX3, which performs *ad hoc* data correction following standard procedures.[29] The parameters $Q_{\mathrm{max,inst}}$, $Q_{\mathrm{max}}$, $Q_{\mathrm{min}}$, and $r_{\mathrm{poly}}$ used to transform each dataset are listed in

**Supplementary Table 2**. Measurements of the empty instrument (*i.e.* air scattering), pure solvent and empty capillary were performed to facilitate accurate background subtraction. We performed small-box, real-space Rietveld refinement on the PDFs using PDFgui.[30] The refined parameters included scale, lattice parameters, atomic displacement parameters, a correlated motion term, and a crystallite-size damping term. Refinement of discrete structures (*i.e.* clusters) was performed using the DiffPy-CMI framework.[31] The refined parameters included scale, a zoom-scale function that accounts for isotropic expansion or contraction of the structure, and atomic displacement parameters.

## Empirical Potential Structure Refinement

Empirical potential structure refinement (EPSR) was used to generate an atomistic model of a cluster in solution (Keggin in water) consistent with the experimental total scattering data. EPSR combines a conventional Monte Carlo simulation, governed initially by a chemically motivated reference potential, with an empirical correction derived iteratively from the discrepancy between the experimental and simulated scattering functions.[32–34]

For these simulations, the X-ray total scattering data for the cluster in solution were reprocessed using the GudrunX programme, using well-documented routines to obtain a quantitative total scattering structure factor.[35] Background subtraction involved the removal of air scattering, thus preserving the signal from both the cluster and solvent. The corrected intensities were subsequently normalised and converted to the reduced structure factor, $F(Q)$, and Fourier transformed over the $Q$-range 1.4–12.0 $Å^{-1}$ to obtain the PDF, $G(r)$. Note that the data processed in GudrunX use a different formalism than PDFgetX3; full conversion details are in Ref. 36.

The simulation comprised 1 Keggin ion, 6 ammonium ions and 666 water molecules, corresponding to the experimental composition of 1.0 M ammonium metatungstate (with respect to [W]). The species were placed within a cubic simulation box of length ~27.5 Å, giving a total number density of 0.1 atoms $Å^{-3}$. Periodic boundary conditions were applied, and simulations were performed at 300 K in the canonical (NVT) ensemble. The Keggin ion was represented by a semi-rigid molecular geometry. To distinguish chemically distinct atomic environments within the cluster, we assigned separate atom types to tungsten (Wk), inward-facing oxygen (Oc), bridging oxygen connecting neighbouring tungsten centres (Ok), and outward-facing terminal oxygen (Okk). Water molecules were represented by oxygen (Ow) and hydrogen (Hw). Ammonium ions were represented by nitrogen (Nam) and hydrogen (Ham). The force field parameters are provided in **Supplementary Table 3**.

The reference potential described atomic interactions using Lennard-Jones and electrostatic terms. The molecular geometry of water was flexible, with an O–H bond length of 0.976 Å and an H–O–H angle of 104.2°. The initial configuration was equilibrated using the reference potential and a reference simulation generated from an ensemble average of several thousand box configurations against which the effects of experimental refinement could be assessed. Following equilibration of the reference model, an empirical potential was introduced to improve agreement between the simulated and experimental scattering functions. We calculated the simulated scattering from the evolving ensemble of atomic configurations using the partial structure factors and their corresponding X-ray scattering weights. Differences between the calculated and experimental scattering functions were used in the iterative refinement process to generate corrections to the empirical potential, which were incorporated into subsequent Monte Carlo sampling. Over time, this procedure progressively modified the intermolecular correlations while retaining the molecular geometry and constraints imposed by the reference model. The quality of the refinement was assessed by comparing the calculated and experimental scattering functions of the fitted $Q$-range, using an $R$-factor.

Following refinement, structural statistics were accumulated from approximately 5,000 configurations sampled over successive model-box configurations. In each configuration, every atom and molecule in the simulation was moved at least once. We obtained equivalent structural statistics from the unrefined reference simulation configurations, allowing direct comparison between the initial and experimentally refined models. The local hydration environment of the Keggin was characterised using site–site radial distribution functions (RDFs), coordination numbers, angular distributions and three-dimensional spatial density functions (SDFs). All structural quantities were calculated from the sampled configurations and averaged over the corresponding ensembles. For SDFs, we accumulated water oxygen (Ow) and hydrogen (Hw) positions on a three-dimensional grid with a voxel step of 0.35 Å within a spherical region extending 8.5 Å from the Keggin centre. This distance was selected to encompass the molecular framework and its immediate hydration environment, as indicated by the RDF analysis. Isosurfaces were generated at three times the respective bulk atomic number densities to highlight regions of preferential solvent occupancy.

**Sample Preparation**

Silicon powder, cerium dioxide ($CeO_2$), and ammonium metatungstate hydrate ($(NH_4)_6[H_2W_{12}O_{40}]$. $x H_2O$) were purchased from Sigma Aldrich. All chemicals were used as received. Ammonium metatungstate was dissolved in water to produce a 1.0 M solution, with respect to [W]. Silicon powder (APS 1–5 μm, Thermo Scientific) was mixed with water to produce a 0.5 M dispersion for measurement in the liquid jet. Ferrihydrite was produced using the solid-state synthesis described in Ref. 37. Nanocrystals of $CsPbBr_3$ (~7 nm diameter) were synthesised following Ref. 38. We then suspended the nanocrystals in toluene for delivery in the liquid jet, in recycling mode.

## Results and Discussion

Our experimental setup was applied to a range of sample states, including highly crystalline powders ($CeO_2$ and Si) and an amorphous/nanocrystalline solid (ferrihydrite), in a capillary geometry. We also employed liquid jet delivery to measure solvents (water and toluene), liquid dispersions (Si in water), photoexcitation of lead halide perovskite nanoparticles ($CsPbBr_3$ in toluene), and clusters in solution (aqueous tungsten-based Keggin anion, $[H_2W_{12}O_{40}]^{6-}$).

### Crystalline Standards: $CeO_2$ and Si

Total scattering data from fused silica capillaries (0.3 mm outer diameter) loaded with either highly crystalline $CeO_2$ or Si powder were measured using a one-pulse-per-train acquisition, with 2,000 pulses averaged. The data spanned a $2\theta$ range of 4.5–85° ($Q$ range of 0.95–16.4 $Å^{-1}$). The $CeO_2$ scattering pattern exhibited sharp diffraction peaks, a low background, and a high signal-to-noise ratio, even at the highest angles (**Fig. 2a**). For Si, sharp peaks and a low background were also observed (**Fig. 2b**).

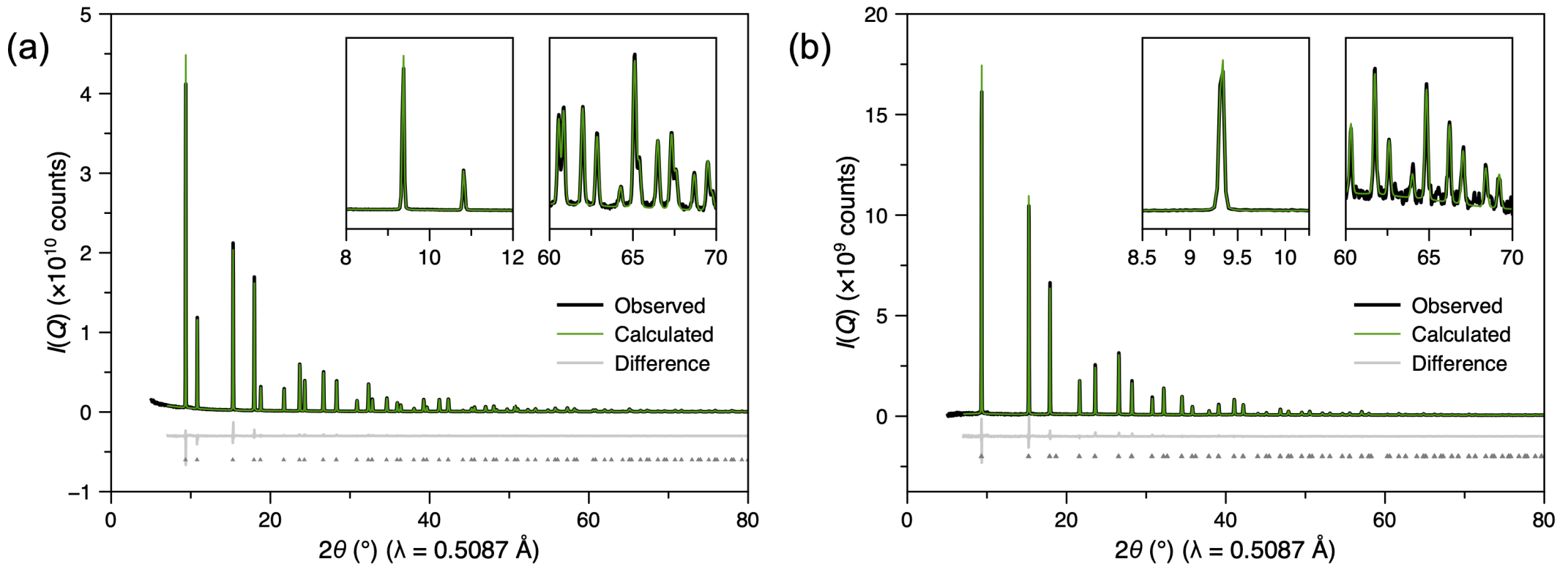

**Figure 2** Rietveld fit to the average total scattering data obtained from **(a)** $CeO_2$ and **(b)** Si. Experimental data (black), calculated data (green), residual (grey), and *hkl* reflections (tick marks). Insets show magnified views of the low- and high-$Q$ regions.

Rietveld refinement against the $CeO_2$ data, with a fixed lattice parameter of 5.411526 Å and a 4th-order $2\theta$ polynomial correction, yielded a $B_{iso}$ of 0.307(4) $Å^2$ for Ce and 0.43(3) $Å^2$ for O ($R_{wp}$ 0.0684). These values match the literature values for $B_{iso}$ (Ce) 0.29 $Å^2$ and $B_{iso}$ (O) 0.43 $Å^2$ and are slightly improved over our previous work (*cf.* 0.253 and 0.48 $Å^2$, respectively).[39] Similar refinement of the Si data, with a fixed lattice parameter of 5.43123 Å, yielded a $B_{iso}$ (Si) of 0.29(1) $Å^2$ ($R_{wp}$ 0.0814). This $B_{iso}$ (Si) value is similar to our previous work (0.328 $Å^2$), but lower than the literature value of 0.47 $Å^2$.[40]

To determine the diffractometer's characteristics, we extracted the instrument resolution function (IRF) as a function of $2\theta$ (**Fig. 3**). The FWHM increased linearly with the scattering angle, varying from about 0.05° to 0.25° over the range 10 to 80°. At the highest angles, a larger spread in FWHM was observed, likely arising from the low intensity of the Bragg peaks beyond 60°. We note the qualitative similarity between the IRF obtained here and that at HED in our previous experiment.[16]

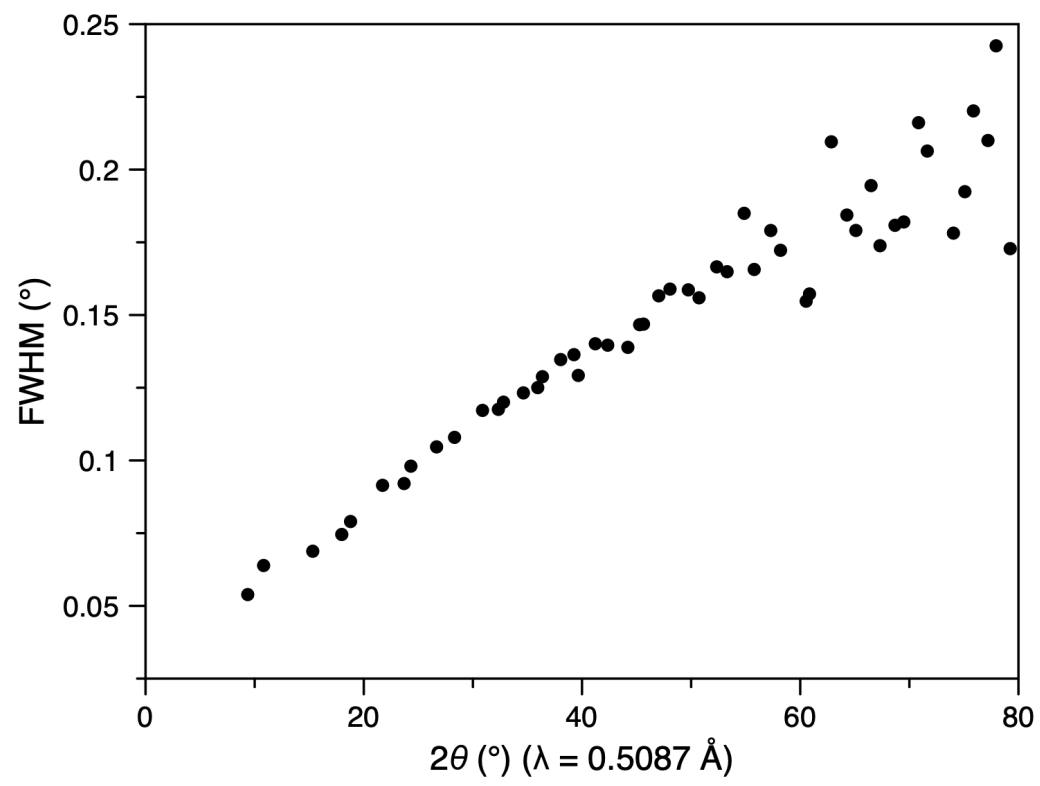

**Figure 3** Bragg peak full width at half-maximum (FWHM) obtained from averaged total scattering data for $CeO_2$ in a capillary, as a function of scattering angle.

We then transformed the total scattering data into PDFs using PDFgetX3 over the $Q$-range of 1–16 Å$^{-1}$.[29] Both PDFs ($CeO_2$ and Si) exhibited sharp peaks across the 0 to 100 Å range, with minimal damping. Small-box Rietveld refinement of $CeO_2$ yielded a good fit ($R_w$ = 0.27), with a lattice parameter of 5.4078(1) Å and $B_{iso}$ (Ce) and $B_{iso}$ (O) values of 0.33(2) and 1.3(1) Å$^2$, respectively (**Fig. 4a & Supplementary Table 4**). The atomic displacement parameters were above (Ce, 0.165 Å$^2$) and below (O, 1.53 Å$^2$) those obtained in our previous demonstration of high-quality PDFs at the HED beamline, while $B_{iso}$ (O) was significantly larger than the value obtained from the reciprocal space data, which is a commonly observed issue in the literature.[41] We determined the instrumental parameters $Q_{damp}$ and $Q_{broad}$ to be 0.0094(7) Å$^{-1}$ and 0.011(1) Å$^{-1}$, respectively. These parameters were marginally higher than those obtained at HED (*cf.* 0.006 and 0.009, respectively). The difference curve appeared randomly distributed, though the amplitude increased slightly at the highest $r$ values, which may reflect the limitations of the instrument's relatively simple $Q_{damp}$ and $Q_{broad}$ parametrisation.[41] We note that the $CeO_2$ data here used only 2% of the XFEL beam intensity, compared with 10% in our HED experiment; this may explain the slight degradation in quality.

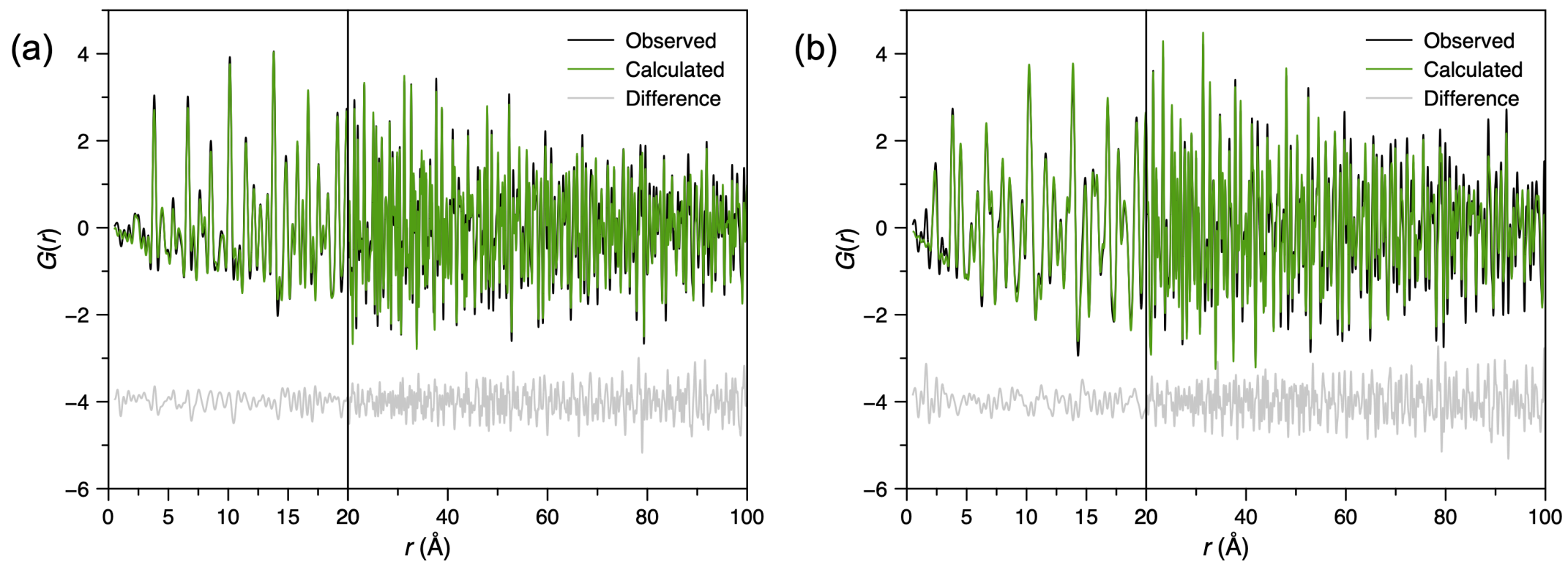

**Figure 4** Small-box Rietveld fits to averaged PDF data from **(a)** $CeO_2$ and **(b)** Si. The vertical line at $r$ = 20 Å indicates a change in the horizontal scale; the vertical scale remains unchanged.

Fitting to the Si PDF data yielded a good fit ($R_w$ = 0.28), with a lattice parameter of 5.4257(1) Å and $B_{iso}$ (Si) of 0.64(1) Å$^2$, slightly higher than our HED result (0.5 Å$^2$) (**Fig. 4b & Supplementary Table 5**). The difference curve was similar to that observed for $CeO_2$, with a slightly increasing amplitude at high $r$. However, there also appeared to be an additional signal in the difference curve at the lowest $r$ values. This likely stems from the lower signal-to-noise observed in the scattering data at the highest angles for this sample.

## Amorphous/Nanocrystalline Solid: Ferrihydrite

Total scattering and PDF measurements are essential for studying the structure of amorphous and nanocrystalline solids. Ferrihydrite is one such material whose structure remains highly debated and has been variously described as amorphous, nanocrystalline, or nanocomposite (small nanocrystallites embedded within an amorphous matrix).[42,43] Despite this uncertainty, its scattering pattern is well documented.[44] Ferrihydrite, more specifically 2-line ferrihydrite, is characterised by two broad diffraction peaks centred around $Q$ = 2.48 and 4.23 Å$^{-1}$, which were both clearly observed in our data (**Fig. 5a & Supplementary Fig. 2**). Furthermore, comparison with total scattering data collected using synchrotron radiation shows excellent agreement; the slight increase in noise at low-$Q$ can be attributed to the subtraction of the rapidly changing background scattering signal from air (**Supplementary Fig. 3**). The XFEL total scattering data were then transformed into the PDF using PDFgetX3 over a $Q$-range of 1.6–16 Å$^{-1}$.[29] The corresponding PDF exhibited relatively sharp peaks at low $r$ (< 10 Å), which broadened rapidly and became featureless above 20 Å (**Fig. 5b**).

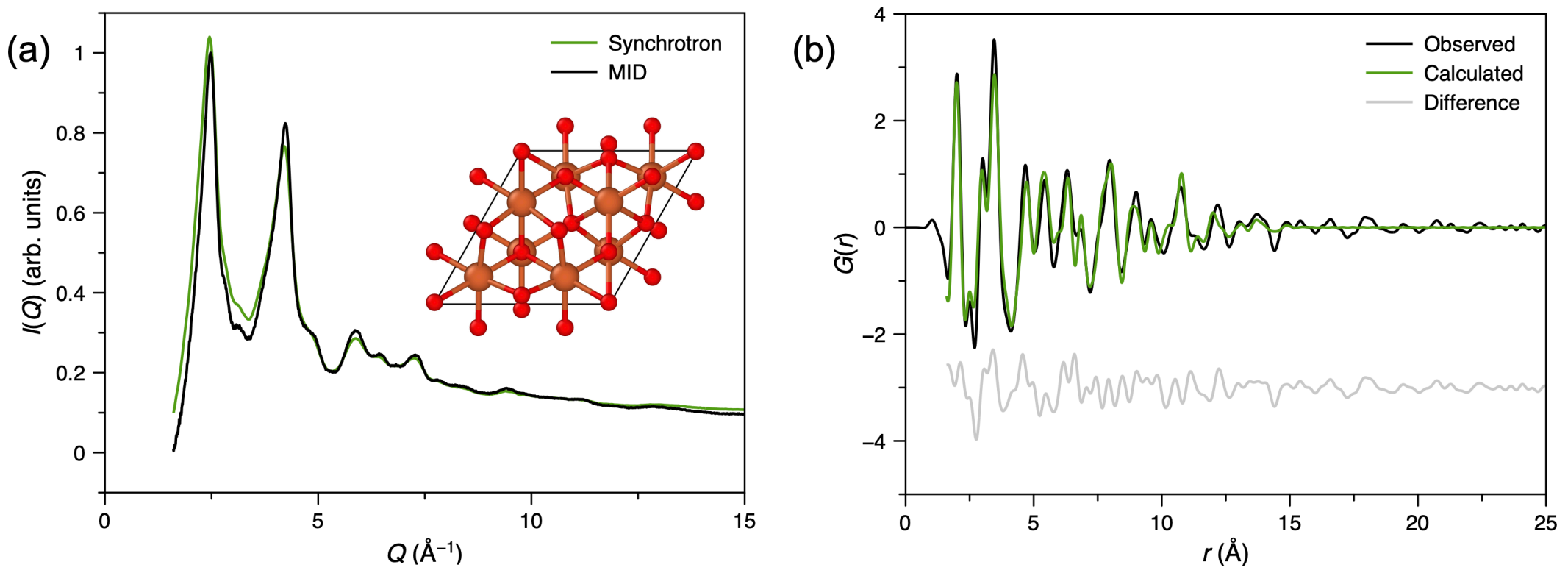

**Figure 5 (a)** Average, background-subtracted, total scattering data obtained from 2-line ferrihydrite (black), compared to synchrotron data (DanMAX, MAX IV) (green); rescaled for comparison. Inset shows the proposed atomic structure by Michel *et al.* (Fe, orange and O, red). **(b)** Small-box Rietveld fit to the corresponding PDF data.

Although the exact structure of ferrihydrite remains unresolved, Michel's 2007 model remains a strong reference point.[42] Using this structure, we performed small-box Rietveld refinement and obtained a fit consistent with previous synchrotron-based measurements ($R_w$ = 0.39) (**Supplementary Table 6**).[37] We determined the *a* and *c* lattice parameters to be 5.95(1) and 9.01(3) Å, respectively, and the spherical crystallite size term was refined to 16.4(7) Å. The $B_{iso}$ values for Fe and O were 0.55(8) and 1.3(3) Å$^2$, respectively. The slight mismatch in relative peak intensities between $r$ = 4.7 and 6.5 Å is common in the literature and indicates variations in Fe site occupancies, which were fixed here to the values reported by Michel *et al.*[42]

## Solvent Scattering: Water

Having established the setup's performance using solid samples in capillary geometry, we next turn to liquid-jet delivery, a key requirement for many pump–probe total-scattering experiments.[6,21,45] Continuously refreshed jets allow each XFEL pulse to probe a new sample volume and are typically compatible with optical excitation, making them well suited to time-resolved studies of solution-phase dynamics. However, they also introduce significant challenges, including weak sample scattering, jet-position reproducibility, and large air/solvent backgrounds.

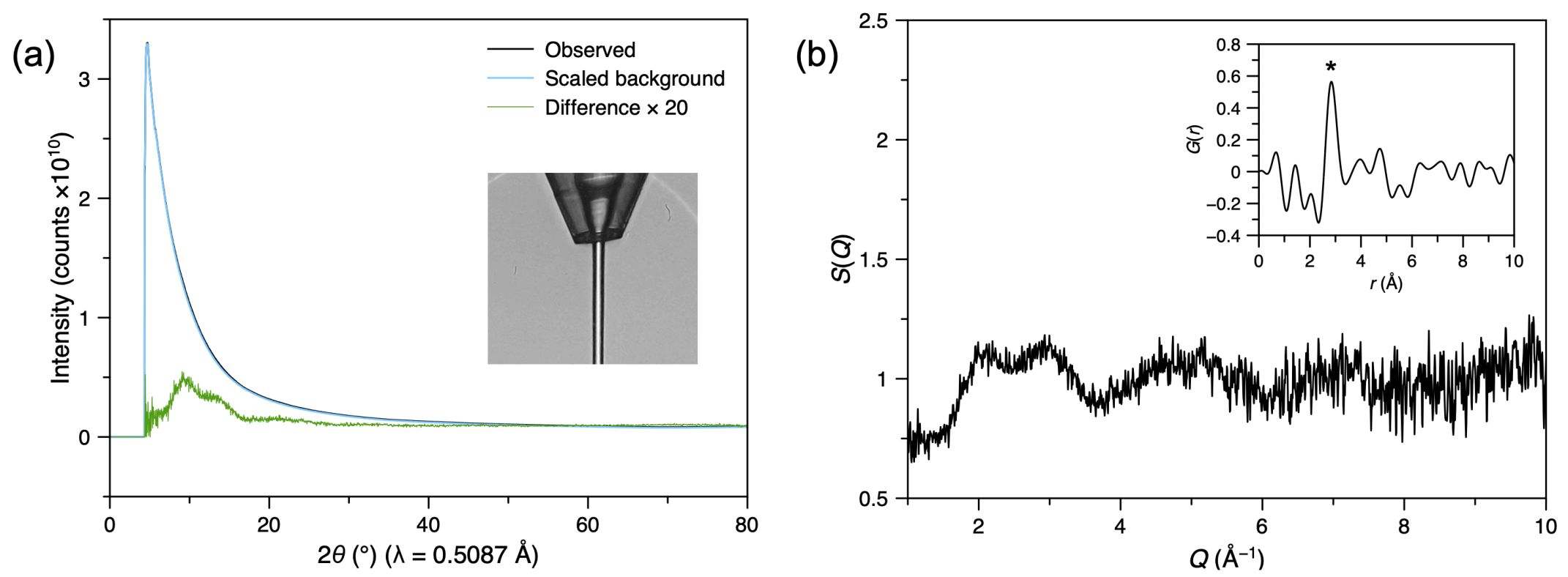

**Figure 6** **(a)** Average total scattering data obtained from water delivered in a liquid jet, compared to the air scattering background, and the resulting water scattering. Inset shows the liquid jet delivered from the nozzle. **(b)** The corresponding average structure factor obtained from water. Inset shows the PDF of water, with the O–O correlation marked with an asterisk.

We first assess the liquid-jet geometry using pure water as a simple benchmark. Bauer *et al.* previously demonstrated single-pulse PDFs of liquid water at MID using a liquid jet under vacuum, with a maximum $Q$-range of 6.38 $Å^{-1}$.[46] Here, we extend the accessible $Q$-range to 12 $Å^{-1}$, despite the substantial air-scattering background. Given the jet's narrow diameter and the comparatively large path length for air scattering, it is remarkable that even a weakly scattering liquid such as water can be detected (**Fig. 6a**). Despite 97.9% of the observed signal arising from air scattering, the signal from water was clearly visible. The structure factor, $S(Q)$, displayed the two characteristic features of water at $Q = 2.1$ and 2.9 $Å^{-1}$ (**Fig. 6b**).[47,48] Beyond this, broad features were observed but became increasingly obscured by noise at higher $Q$. Nevertheless, the PDF obtained by Fourier transforming the data over a $Q$-range of 1.0–12 $Å^{-1}$ exhibited a clear peak at 2.80 Å arising from the O–O correlations within water (**Fig. 6b inset**).

## Liquid Dispersion: Si in water

More commonly, the sample of interest is a solute dissolved or suspended in a solvent, which is then delivered *via* a liquid jet. To test the setup's sensitivity, we first prepared a liquid suspension of Si (1–5 µm crystal size) in water (0.5 M). Due to challenges in subtracting contributions from both water and air to the background, we subtracted only the air scattering, which accounted for 86.3% of the observed signal (**Fig. 7a**). The resulting difference curve showed clear Bragg peaks from Si, superimposed on a diffuse background signal from water (particularly at low scattering angles, where the key features of water scattering are prominent) (**Fig. 7a inset**). These Bragg peaks were visible up to around $Q = 8.7$ $Å^{-1}$, after which they became indistinguishable from the noise and solvent scattering. Because direct subtraction of the measured water background was unreliable, we fitted a cubic smoothing spline to the data, excluding Bragg peaks, and subtracted it to restore a flat baseline at $S(Q) = 1$ before Fourier transformation (**Supplementary Fig. 4**).

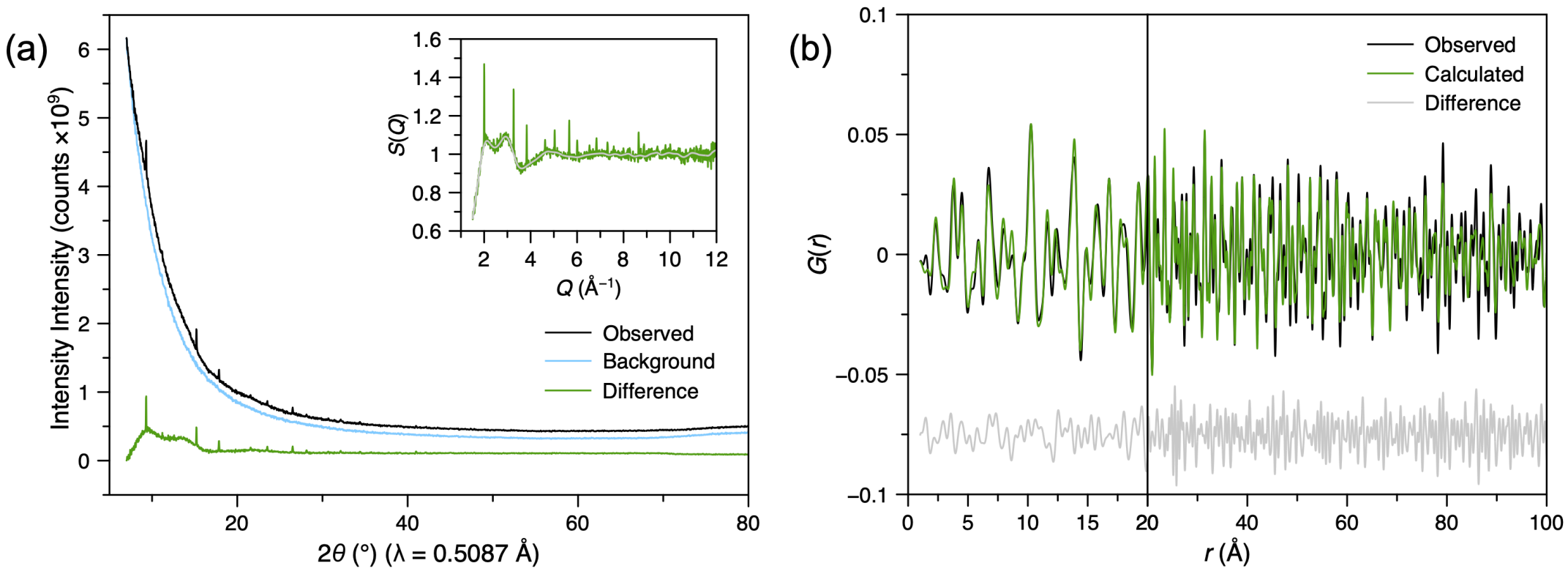

**Figure 7 (a)** Average total scattering data obtained from liquid dispersion of Si in water delivered in a liquid jet, compared to the air scattering, and resulting Si scattering superimposed on the remnant water scattering signal. The inset shows the S(Q) (green) and the cubic smoothing spline (grey) used to correct for the water contribution. **(b)** Corresponding small-box Rietveld refinement of the PDF obtained from Si in water. The vertical line at $r$ = 20 Å indicates a change in the horizontal scale; the vertical scale remains unchanged.

The PDF obtained from the Fourier transform of the data in the $Q$-range of 1.5–12 Å$^{-1}$ displayed clear peaks from Si to high $r$. Small-box Rietveld refinement to the PDF up to 100 Å yielded a good fit ($R_w$ = 0.38) with a refined lattice parameter of 5.44(2) Å and $B_{iso}$ (Si) of 1.4 Å$^2$ (**Fig. 7b & Supplementary Table 7**). The lattice parameter agreed well with the PDF data from Si powder in a capillary (*cf.* 5.4257(1) Å). However, the uncertainty in $B_{iso}$ (Si) was larger than the refined value, suggesting that reliably determining atomic displacement parameters from liquid-jet samples may be challenging. The Si suspension measured here had a relatively low concentration of 0.5 M (~14 g L$^{-1}$). Increasing the Si concentration while maintaining liquid-jet compatibility, along with minimising air-scattering contributions through improved experimental geometry, could substantially improve the signal-to-noise ratio at high $Q$ and, consequently, yield more robust PDFs.

## Photoexcitation in Solution: Lead Halide Perovskite Nanocrystals

We next consider a substantially more demanding nanoscale system: $CsPbBr_3$ nanocrystals (~7 nm diameter) suspended in toluene and delivered by the liquid jet. Compared with the larger Si particles discussed above, their smaller size produces intrinsically broadened diffraction features and diffuse scattering below and between peaks, making structural changes harder to resolve against the solvent and air background, despite the much stronger scattering power. This sample also enables our first assessment of pump–probe total-scattering measurements at MID: above-gap optical excitation initiates a transient structural response in the nanocrystals, which femtosecond XFEL pulses can then probe at defined pump–probe delays. $CsPbBr_3$ nanocrystals therefore bridge the static liquid-jet benchmarks described above and the longer-term goal of high-quality, time-resolved measurements of photoinduced structural dynamics in solution. In this regard, metal halide perovskites are known to exhibit ultrafast localised lattice distortions (polaron formation) upon photoexcitation.[9] Ultrafast optical-pump–X-ray-probe experiments on monodisperse nanocrystals are key for investigating these phenomena, their structural aspects, and photophysical implications.[24]

Nanocrystals of $CsPbBr_3$ were first synthesised and characterised by TEM (**Supplementary Fig. 5**). The nanocrystals were then suspended in toluene and delivered *via* the liquid jet system. We performed optical-pump–X-ray-probe experiments using a 400 nm pump laser with a +5 ps delay and pump fluences of 0.82, 1.97, 3.94, and 5.77 mJ $cm^{-2}$. Separate measurements of the toluene jet and air scattering were collected to correct for background scattering, which accounted for ~96% of the total signal (**Fig. 8a**).

We first investigated the effects of above-gap photoexcitation by analysing the normalised differential scattering signals $(I - I_0)/I_0$, where $I$ and $I_0$ denote the scattering intensities measured at +5 and −5 ps delay relative to the pump pulse, respectively. The resulting differential signals for the background-corrected data at increasing excitation fluence exhibit progressive appearance of negative features toward the high-$Q$ side and the emergence of positive features on the low-$Q$ side of the main peaks (**Fig. 8b**). This result is particularly notable given the noisy background-subtracted signals, an issue further compounded by the experimental conditions required to maximise photoexcitation of the nanocrystals. Between fluences, we observed an increase in the differential signal from 0.82 to 3.94 mJ $cm^{-2}$, followed by a decrease in signal and slight broadening at the highest fluence.

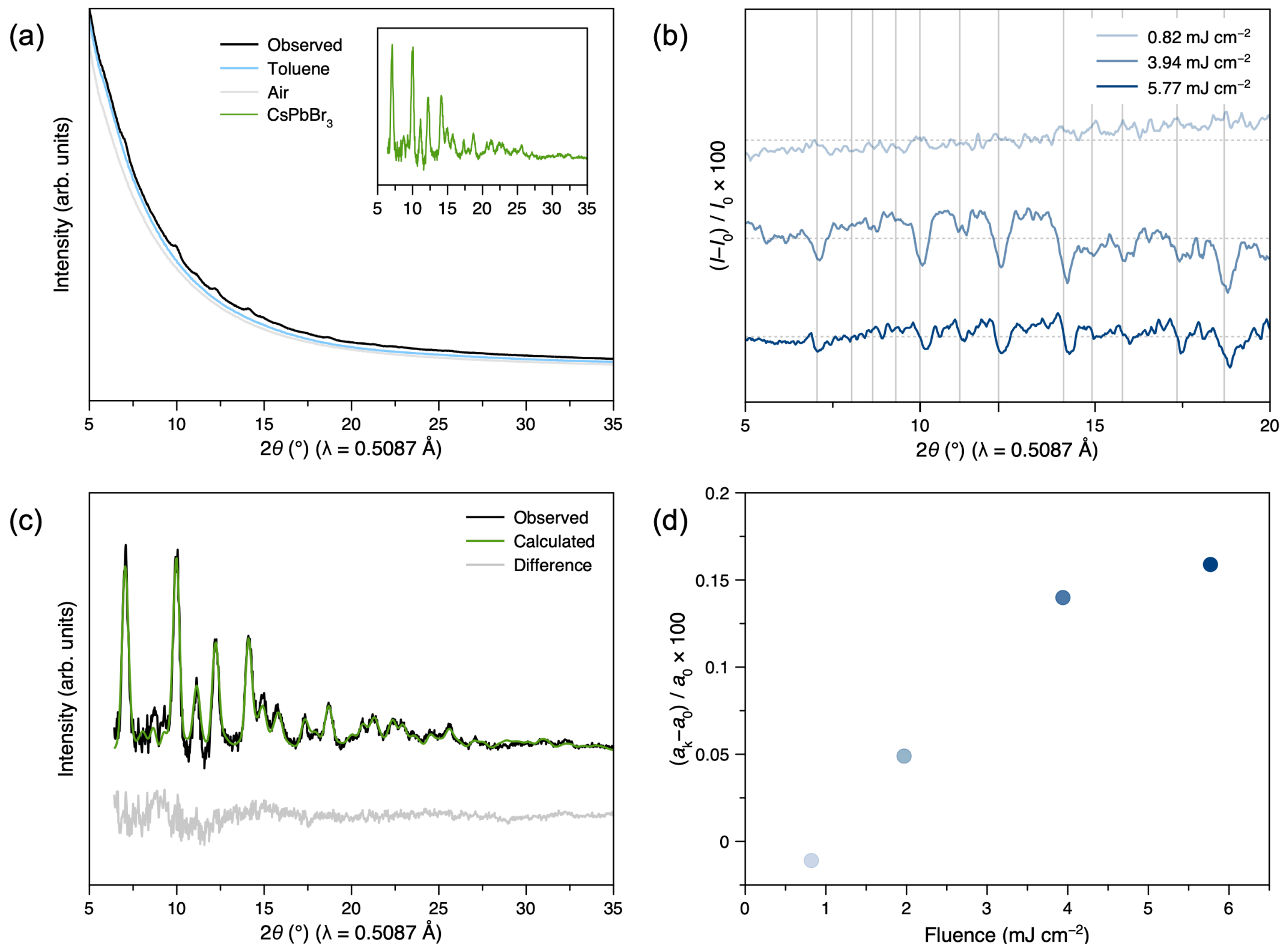

**Figure 8** **(a)** Average total scattering data from $CsPbBr_3$ colloidal nanocrystals in toluene delivered in the liquid jet, compared to the toluene and air scattering, and the derived $CsPbBr_3$ signal (inset). **(b)** Normalised differential scattering signals after background subtraction, measured at a delay time of +5 ps. Vertical lines indicate positions of the main Bragg peaks for $CsPbBr_3$. The curves are vertically offset for clarity, with dashed horizontal lines indicating the zero level for each dataset. **(c)** Average, background-subtracted data collected at the highest pump fluence (5.77 mJ cm$^{-2}$), with the model constructed using the DSE approach. **(d)** Relative isotropic strain, expressed as the difference between the equivalent cubic lattice parameter of CsPbBr3 refined from the photoexcited (+5 ps, $a_k = (abc/4)^{1/3}$) and the reference (−5 ps, $a_0 = 5.842(2)$ Å) datasets as a function of fluence.

To quantitatively investigate the structural response of $CsPbBr_3$ nanocrystals upon photoexcitation, we analysed the reciprocal-space data using the Debye scattering approach. A reference structural model was derived from high-resolution X-ray total scattering data collected at the ID22 beamline of the European Synchrotron Radiation Facility from a suspension of the same $CsPbBr_3$ nanocrystals. The Rietveld-refined lattice parameters from these data using the orthorhombic *Pbnm* crystal structure were $a$ = 8.161(5), $b$ = 8.277(4) and $c$ = 11.782(5) Å, while the equatorial and axial Pb–Br–Pb bond angles were 158° and 166°, respectively.[49,50]

Starting from this set of parameters, we generated atomistic models of nanocrystals with increasing size along two orthogonal growth directions, namely $D_{ab}$ (lying in the **ab**-plane) and $L_c$ (parallel to the **c**-axis), following the standard DebUsSy strategy for a bivariate population of nanocrystals with a cuboidal shape.[27] We used these models to compute interatomic distances between atom pairs in the nanocrystals, which are needed to generate the corresponding X-ray total scattering patterns calculated using the Debye scattering equation (DSE). We used this structural/morphological model to perform DSE analysis of both the ESRF data and the present XFEL data (**Supplementary Fig. 6**). **Figure 8c** shows the best fit for the pumped state at the highest fluence (5.77 mJ $cm^{-2}$) after background subtraction. During model optimisation, we refined an isotropic strain parameter (averaged over the randomly oriented ensemble of NCs, uniformly expanding or contracting the interatomic distances) and the atomic displacement parameters (in the form of isotropic Debye-Waller factors, $B_{iso}$) for all atomic species against the experimental data for the non-excited ($I_0$, equilibrium state) and each excited sample. We constrained the average nanocrystal size and size distribution to values from the high-resolution synchrotron measurements, according to a bivariate lognormal distribution. The refined parameters are summarised in **Supplementary Table 8**. We extracted a small, progressive lattice expansion, up to 0.17% relative to the equilibrium state, as pump fluence increased (**Fig. 8d**). We also note that the refined atomic displacement parameters do not indicate a systematic increase in dynamic disorder upon photoexcitation. These very preliminary findings appear consistent with similar observations reported for bulk perovskites, supporting the suitability of our experiments.[9]

In summary, our measurements on $CsPbBr_3$ nanocrystals show that pump–probe data can also be obtained from challenging samples. Our observations show a fluence-dependent increase in isotropic strain, without a coupled, systematic increase in the Debye-Waller factors. These preliminary results are promising for future studies of the ultrafast photoexcited dynamics occurring in perovskite nanocrystals.

**Cluster in Solution: Aqueous Keggin**

As a final demonstration of the liquid jet, we measured an aqueous solution of ammonium metatungstate, in which the Keggin ion ($[H_2W_{12}O_{40}]^{6-}$) is released. In our previous work, the Keggin cluster (~8 Å diameter) served as the most rigorous test of the experimental setup, and it is therefore a valuable benchmark system. We prepared a 1.0 M aqueous solution of ammonium metatungstate (relative to [W]) and delivered it *via* the liquid jet. Background subtraction for clusters in solution is challenging, even with the highest-quality data; here, with such a narrow liquid jet and extensive air scattering, the challenge is much greater. Nonetheless, we used a combined air-and-water scattering measurement as the background, which accounted for 86.6% of the observed signal, similar to the Si suspension (**Fig. 9a**). As a cluster in solution, this sample is not expected to show Bragg peaks in its scattering data. However, due to the high scattering power of tungsten, the diffuse scattering signal from the cluster was relatively low-noise, even at the highest scattering angles.

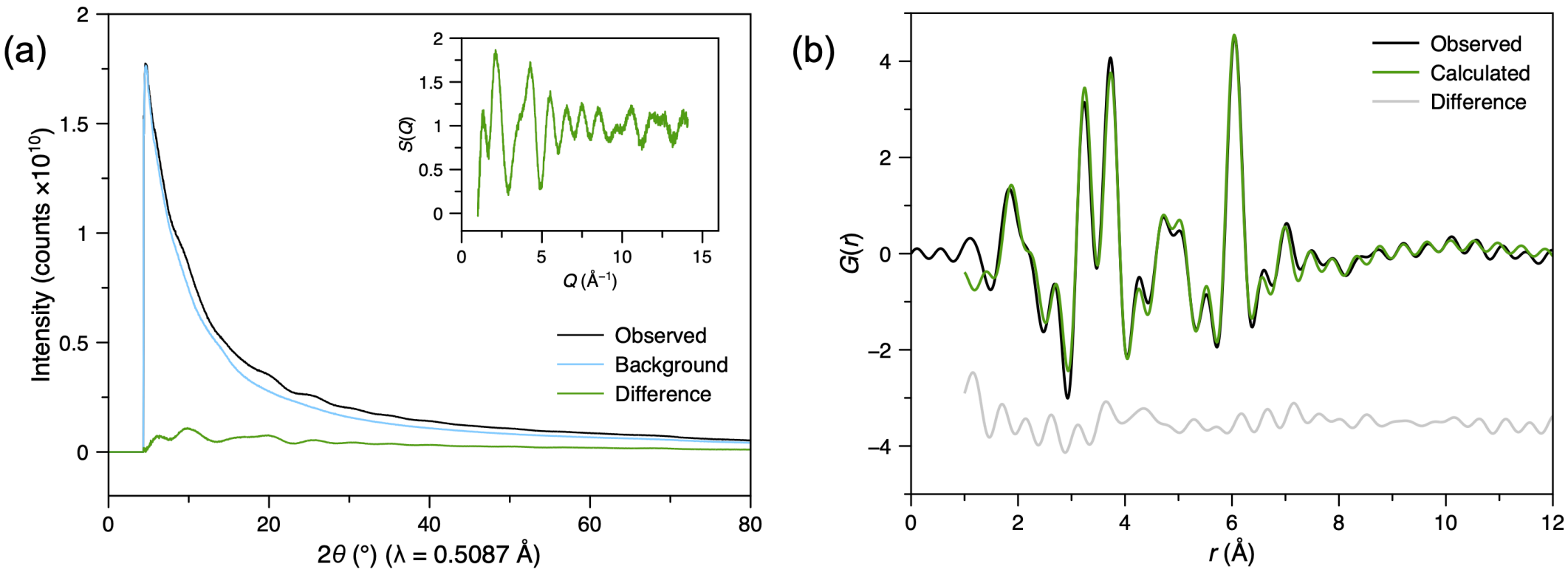

**Figure 9 (a)** Average total scattering data obtained from the Keggin in water delivered in a liquid jet, compared to the background scattering (air and water), and the resulting Keggin signal. Inset shows the corresponding $S(Q)$. **(b)** Corresponding structure refinement against the average PDF obtained from Keggin.

The PDF obtained from the Fourier transform of the data in the $Q$-range of 1–14.1 $Å^{-1}$ contained several sharp peaks at low $r$, before terminating around 8 Å (the approximate size of the cluster) (**Fig. 9b**). Key peaks in the PDF were observed at $r$ = 1.84 Å (W–O), 3.24 Å (W–W), 3.73 Å (W–W), and 6.05 Å (W–W). We refined the Keggin cluster structure against the PDF using cluster modelling in the DiffPy-CMI package.[31] Here, the calculated PDF was derived from the discrete model *via* the $I(Q)$ computed from the Debye scattering equation. To refine the model against the data, we refined an overall scale factor, independent isotropic atomic displacement parameters for W and O, and a zoom-scale parameter that isotropically expands or contracts the structure while maintaining relative atomic positions. We obtained an excellent fit, comparable to our previous measurements at HED ($R_w$ = 0.20), with a zoom-scale parameter close to unity (1.007(8)) (**Supplementary Table 9**). The atomic displacement parameter $B_{iso}$ (W) was determined to be 0.23(4) $Å^2$ (*cf.* 0.22 $Å^2$ from our previous work); however, we could not reliably determine $B_{iso}$ (O). Taken together with the results from the Si liquid suspension, this suggests that PDFs of liquid-jet-delivered samples can yield reliable atomic displacement parameters for heavier elements, such as W, but that this approach is challenging for lighter elements, such as Si and O.

In solution, not only does the observed total scattering signal contain information regarding the solute (Keggin ion) and solvent (water) interactions, but also the interaction between the solute and the solvent. Probing these interactions is important for understanding solvation structure, which can strongly affect reactivity and solution-state reaction dynamics. Previous work using XFEL scattering measurements has typically focused on elucidating these interactions using relatively low-$Q$ data (*i.e.*, time-resolved X-ray solution scattering methods).[45,51–55] By extending our scattering to higher $Q$, we can gain greater sensitivity to local solvation structures and allow for more sophisticated analysis methods.

EPSR was used to obtain an atomistic description of the hydration environment surrounding the Keggin ion, constrained by the experimental X-ray total scattering data. EPSR has been widely applied to synchrotron X-ray and neutron scattering to study hydrogen-bond networks in liquids, ion solvation and association in aqueous solutions, and the local atomic structure of amorphous materials, but it has not yet been used to analyse XFEL total scattering data.[56–58] A detailed discussion of this method is provided in Ref. 32 and in the **Methods** section.

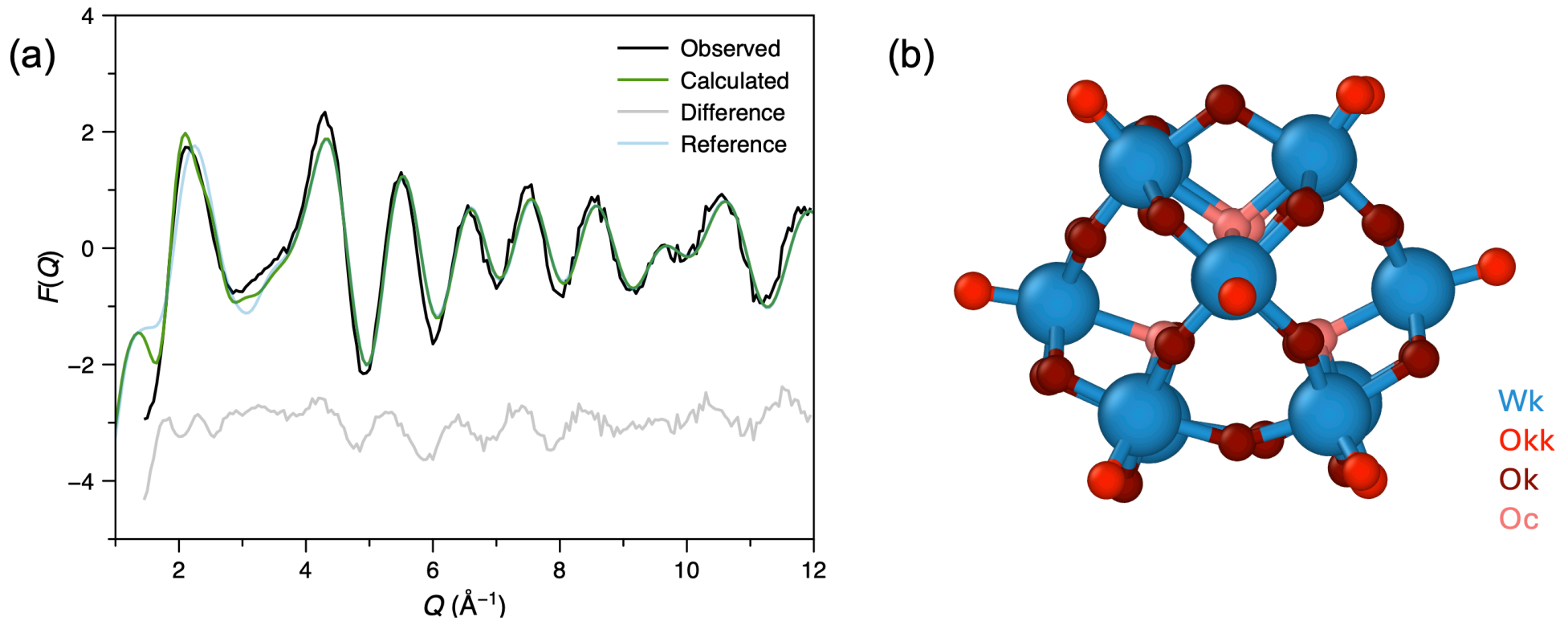

**Figure 10 (a)** EPSR fit to the $F(Q)$ data from the Keggin in water. The simulation from the unrefined reference potential is shown for comparison. **(b)** Atomic structure of the Keggin ion with the four different atomic sites labelled.

The reference simulation, performed using the initial intermolecular potential without empirical correction, reproduced the principal features of the experimental scattering, indicating that the reference potential already provides a reasonable description of the solution structure (**Fig. 10a**). Subsequent empirical potential refinement improved the agreement, particularly around the first scattering maximum at approximately $Q$ = 2–3 Å$^{-1}$. The $R$-factor decreased from 0.12 for the reference model to 0.08 after refinement. The good agreement obtained with the reference potential alone suggests that only modest empirical corrections were required to optimise the structural model. Some discrepancies remain, particularly below $Q$ = 2 Å$^{-1}$, which are typical of this type of experiment and reflect difficulties in treating the Compton scattering background.[59] The remaining systematic oscillations in the difference curve are consistent with a slight mismatch in peak positions, potentially arising from the approximately 1% expansion of the Keggin framework identified by independent real-space PDF refinement but not incorporated into the EPSR model. The refined configurations were subsequently analysed to characterise the Keggin hydration environment.

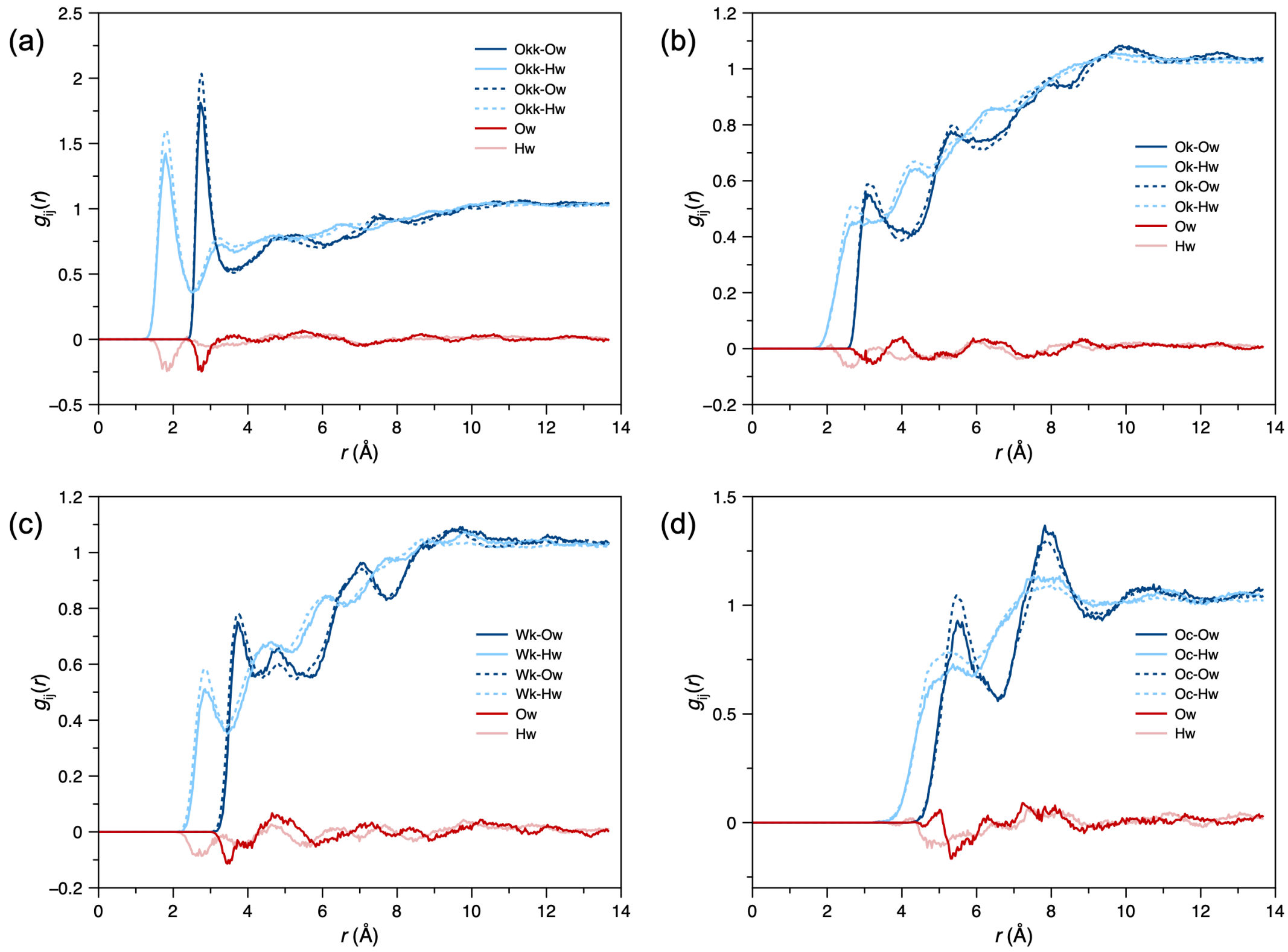

**Figure 11** Partial radial distribution functions, $g_{ij}(r)$, describing hydration around **(a)** terminal oxygen (Okk), **(b)** bridging oxygen (Ok), **(c)** tungsten (Wk), and **(d)** inward-facing oxygen (Oc) sites of the Keggin ion. Dark and light blue curves correspond to correlations with water oxygen (Ow) and hydrogen (Hw), respectively. Dashed and solid lines show distributions before and after EPSR refinement, respectively. Red and pink curves show the corresponding changes for Ow and Hw correlations, respectively.

The hydration environment of the Keggin ion was investigated through partial radial distribution functions, $g_{ij}(r)$, calculated between distinct Keggin sites (**Fig. 10b**) and water oxygen (Ow) and hydrogen (Hw) atoms. The pertinent RDFs are shown in **Fig. 11**, with the corresponding reference and refined distributions compared to establish the effect of experimental refinement. The four distinct Keggin sites show markedly different solvent correlations, reflecting the cluster surface's heterogeneous chemical and geometric character.

The most pronounced short-range hydration features are associated with the outward-facing terminal oxygen atoms (Okk; **Fig. 11a**). The Okk–Hw RDF exhibits a sharp first maximum at approximately 1.8 Å, while the corresponding Okk–Ow distribution displays a maximum at 2.8 Å. The separation between these maxima is consistent with water molecules preferentially orienting the hydrogen in their O–H bond towards the terminal oxygen, as expected for hydrogen-bond donation from water to the Keggin surface. The first hydration-shell coordination number, obtained by integrating the Okk–Ow RDF to 3.2 Å, was ~2.5 water oxygen atoms per terminal oxygen site. Comparison between the reference and refined distributions reveals a reduction in the intensity of the first Okk–Ow maximum, from $g(r)$ = 2.0 to 1.8, accompanied by a corresponding reduction in the first Okk–Hw peak from 1.6 to 1.4. The peak positions remain largely unchanged, indicating that refinement primarily modifies the water population around the terminal oxygen sites.

The bridging oxygen atoms (Ok) exhibit a less sharply defined hydration environment than the terminal oxygens (**Fig. 11b**). The Ok–Hw and Ok–Ow distributions have broad first-neighbour features around 2.0–2.5 Å and 3–3.5 Å, respectively. The broader, less intense correlations suggest that water molecules around bridging oxygens occupy a wider range of configurations. This behaviour is consistent with the partial geometric shielding of bridging oxygen sites by neighbouring W–O units. The reference and refined Ok-centred RDFs display similar overall features, although refinement produces a decrease in intensity and a noticeable broadening of the Ok–Hw correlations.

The tungsten-centred RDFs exhibit a more complex sequence of solvent correlations (**Fig. 11c**). The first appreciable Wk–Hw and Wk–Ow correlations occur at approximately 2.8–3.0 Å and 3.7–3.9 Å, respectively, followed by multiple broader features. Unlike the terminal and bridging oxygen correlations, these Wk-centred distributions do not necessarily represent direct intermolecular coordination to tungsten. Rather, they contain contributions from water molecules interacting with multiple oxygen sites distributed around the W–O framework. The longer-distance oscillations therefore reflect, at least in part, the geometry of the Keggin ion and the spatial distribution of its surrounding hydration environment.

The inward-facing oxygen sites (Oc) exhibit a qualitatively different hydration profile (**Fig. 11d**). Both Oc–Ow and Oc–Hw distributions show negligible correlations at short distances, with water density becoming appreciable only beyond approximately 4 Å. This behaviour is consistent with the location of these oxygen atoms within the Keggin framework, where direct access by solvent molecules is sterically restricted. The Oc–Ow distribution exhibits a pronounced maximum at approximately 5.5 Å, accompanied by a corresponding feature in the Oc–Hw distribution. Because the reference oxygen sites lie within the cluster rather than at its outer surface, these longer-range features primarily reflect the radial distribution of water surrounding the complete Keggin structure.

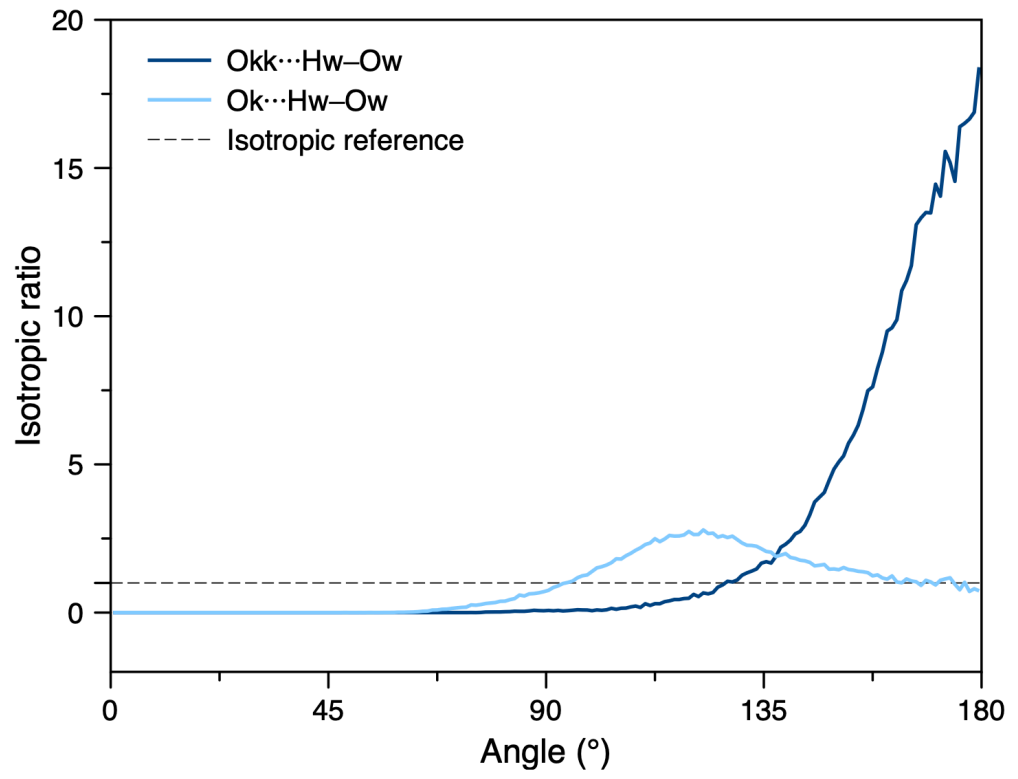

**Figure 12** Angular distributions of Okk···Hw–Ow (dark blue) and Ok···Hw–Ow (light blue) contacts calculated from the EPSR configurations, where Okk and Ok denote terminal and bridging Keggin oxygens, respectively, and Hw and Ow denote water hydrogen and oxygen, respectively. Angles are expressed as dimensionless isotropic ratios: observed angular bin probabilities are divided by those expected for independent, randomly oriented directions. The dashed line at unity marks the isotropic reference. Terminal oxygen contacts favour near-linear geometries, whereas bridging oxygen contacts exhibit a broad preference around 120°.

To probe the molecular geometry of the hydration environment, angular distributions were calculated from the sampled configurations (**Fig. 12**) and expressed as isotropic ratios, which quantify angular preferences relative to independently random orientations. The short Okk–Hw and Okk–Ow distances suggest that water molecules preferentially orient one hydrogen atom towards the terminal oxygen sites. This was investigated by examining the Okk···Hw–Ow and Ok···Hw–Ow angular distributions for water molecules satisfying an Okk···Hw and Ok···Hw distance of up to 2.5 and 3.1 Å, respectively. The Okk distribution rises sharply above ~140°, peaks near 180°, and has an isotropic ratio of ~18. The pronounced enhancement in the 160–180° range indicates a strong preference for near-linear geometries, supporting the assignment of short-range RDF correlations to hydrogen-bonded water configurations.

The corresponding Ok···Hw–Ow distribution exhibits a broad maximum around 120°, with a maximum isotropic ratio of approximately 2.7 and enhancement above the isotropic reference extending over approximately 95–165°. Compared with terminal oxygen sites, the bridging oxygen distribution is less strongly localised and favours substantially bent contacts. This behaviour is consistent with the distinct steric and chemical environments of the two oxygen types, although the angular distributions alone do not establish differences in hydrogen-bond energies or populations.

Together, these results provide a molecular interpretation of the site-specific RDFs. The terminal oxygens favour water molecules donating nearly linear hydrogen bonds, while the bridging oxygens exhibit a broader range of predominantly bent contact geometries. The hydration environment is therefore characterised by chemically distinct local water configurations across the Keggin surface. We note that the orientational distributions are derived from the EPSR configurations and may depend on the adopted water model, reference intermolecular potential and contact-selection criteria. In particular, X-ray scattering provides limited direct sensitivity to hydrogen positions. The angular distributions are therefore interpreted as structural characteristics of the experimentally refined molecular model, rather than independently determined experimental observables.

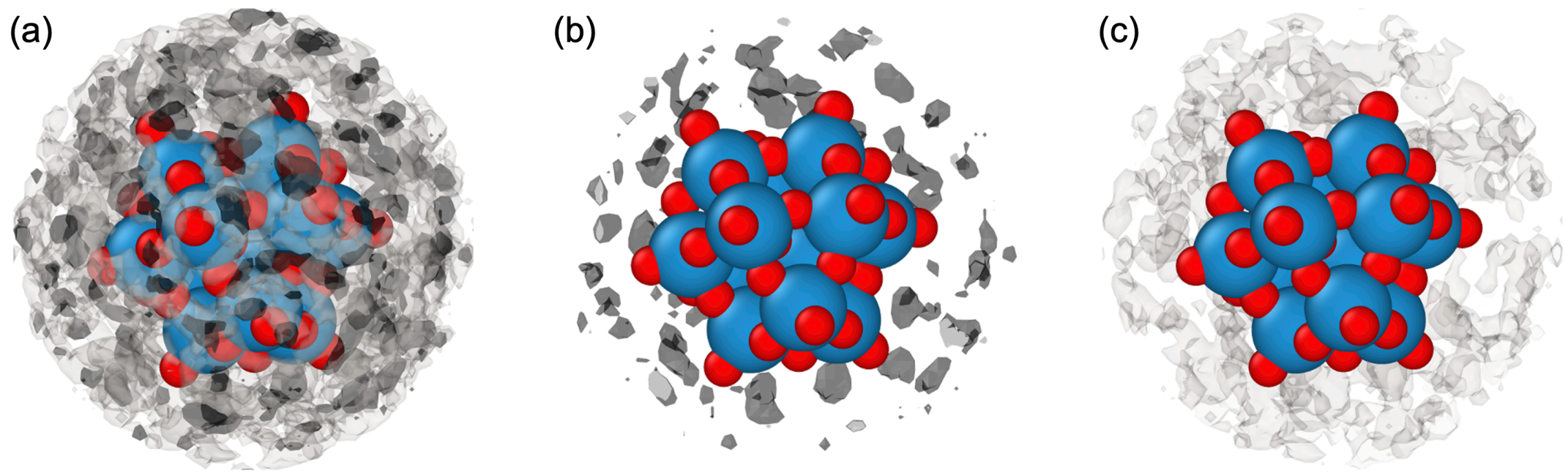

**Figure 13** Three-dimensional spatial density functions (SDFs) of water surrounding the Keggin ion, calculated from the EPSR-refined configurations. **(a)** Combined water-oxygen (Ow, black) and water-hydrogen (Hw, grey) distributions. **(b)** Water-oxygen and **(c)** water-hydrogen distributions shown for the rear hemisphere only, with the complete Keggin structure retained for reference. The Keggin structure is represented by W (blue) and O (red) spheres. The isosurfaces are shown at three times the respective bulk atomic-number densities, and highlight regions of enhanced solvent occupancy relative to the bulk liquid.

To further investigate the spatial organisation of water around the Keggin ion, three-dimensional spatial density functions (SDFs) were calculated from the EPSR-refined configurations (**Fig. 13**). The resulting SDFs reveal a heterogeneous distribution of water around the Keggin surface (**Fig. 13a**). The water-oxygen distribution exhibits comparatively localised density features, whereas the corresponding hydrogen distribution is more diffuse (**Fig. 13b & c**). Unlike the discrete solvation lobes observed for some smaller molecular solutes, the Keggin hydration environment comprises a more complex arrangement of spatially distributed density maxima. This may reflect the relatively high density of potential hydrogen-bonding sites across the compact, oxygen-rich molecular surface, which permits multiple, potentially overlapping solvent configurations. The broader hydrogen distribution is also consistent with a range of accessible water orientations around these sites.

Importantly, this analysis shows that experimentally refined atomistic configurations from XFEL liquid-jet total scattering measurements can be used to construct three-dimensional solvent density maps. Although the SDFs represent ensemble-averaged, model-dependent distributions rather than directly measured spatial densities, their reconstruction illustrates the potential to extend XFEL solution scattering beyond conventional one-dimensional structural analysis toward a more detailed description of solute–solvent organisation.

More broadly, the Keggin measurement provides a stringent benchmark for the liquid-jet configuration. Despite the dominant air–water background, high-$Q$ scattering can be isolated, normalised and converted into a PDF for cluster refinement, and used to probe the surrounding solvation environment *via* EPSR. Future studies combining pump–probe total scattering with atomistic modelling could investigate changes in solvent coordination, hydrogen-bonding geometry, and three-dimensional hydration structure following photoexcitation. Such experiments would extend ultrafast PDF measurements beyond transient intramolecular structures towards the coupled dynamics of molecular systems and their surrounding solvent environments.

## Conclusions and Outlook

Instrument developments like those presented here are essential to extend total scattering and PDF analysis into new experimental regimes. High-quality PDF measurements depend not only on access to sufficiently high $Q$ but also on stable geometry, careful background control, robust detector calibration and sample environments that preserve the structural signal of interest. Building on our previous demonstration of high-quality, ultrafast total scattering and PDF measurements, the present work shows that XFEL total scattering data extending to high $Q$ and PDFs of sufficient quality for meaningful structural analysis can be obtained across a diverse range of samples in a more complex yet practical sample environment.

For crystalline and amorphous solids measured in capillary geometry, the data quality satisfies the requirements for conventional reciprocal- and real-space refinement, including the determination of instrumental parameters and structural models for both ordered and disordered materials. More significantly, the liquid-jet measurements demonstrate that meaningful structural information can be recovered from demanding liquid- and solution-phase samples despite dominant air and solvent backgrounds. Water, Si dispersed in water, and aqueous Keggin clusters show that liquid-jet data can be used to extract PDFs suitable for fitting. Beyond the solute structure itself, EPSR refinement of the Keggin total scattering data enabled atomistic modelling of its hydration environment, yielding site-specific RDFs, angular distributions and three-dimensional solvent density maps. These show that the data are consistent with preferential near-linear hydrogen-bond geometries at terminal oxygen sites and more diffuse, bent water contacts at bridging oxygen sites, demonstrating the potential to characterise site-specific solute–solvent interactions from XFEL data. Pump–probe measurements of $CsPbBr_3$ nanocrystals demonstrated sensitivity to photoinduced structural changes, revealing a fluence-dependent isotropic lattice strain at 5 ps after photoexcitation. Together, these results establish the capability for high-$Q$, time-resolved total scattering and PDF measurements in liquid jets, enabling the investigation of transient structural changes and, through atomistic modelling, the surrounding solvent environment.

Several challenges remain before this approach can be considered routine. Background subtraction is the central limitation, particularly for liquid-jet samples, where air scattering dominates the measured intensity and residual solvent scattering can contaminate the solute signal. Accurate background measurements are inherently difficult because they require reproducing the air, solvent, jet position, jet diameter, and beam–jet overlap with high fidelity between sample and background acquisitions, which is further complicated by fluctuations in intensity and energy distribution of the X-ray beam at FELs. The large air path in the present setup is a major source of background scattering. During the experiment, this was mitigated using upstream shielding and a helium-flushed Kapton tube downstream of the sample, although their positioning was difficult to reproduce between measurements. Future improvements should therefore prioritise more permanent shielding and collimation, along with helium-filled and evacuated flight paths that extend close to the sample. Operating the liquid jet under vacuum would further suppress air scattering, as recent developments show, but it introduces challenges related to rapid evaporation, cooling, and jet stability, potentially limiting its applicability to certain samples.[60,61] Alternatively, a horizontally oriented liquid jet could facilitate more effective shielding and improve the reproducibility of absorption corrections, although gravitational deflection would need to be considered.[62] For weakly scattering or low-Z solutes, the signal-to-background ratio remains challenging, and higher concentrations, larger jet diameters, or improved background suppression may be required. However, for optical-pump/X-ray-probe experiments, the optical penetration depth limits how much the jet diameter can increase. Beyond this depth, additional sample thickness contributes predominantly unexcited scattering, diluting the transient signal. Optimising the jet geometry therefore requires balancing scattering intensity against optical excitation efficiency. Our structural refinements also indicate that atomic displacement parameters, particularly for light elements in liquid-jet measurements, should be interpreted with caution.

Despite these challenges, the outlook is highly positive. The current results show that the key ingredients for high-quality total scattering measurements and PDF analysis are now in place: access to high-$Q$ scattering, sufficient detector performance, compatible sample delivery, and a data-processing workflow that can extract structural information in the presence of high backgrounds. With further optimisation of the liquid-jet environment, background protocols, beam conditioning and alignment reproducibility, these developments will enable high-quality, time-resolved PDF measurements of liquids, molecular solutions, clusters and reacting systems, opening new opportunities to investigate ultrafast structural dynamics in real space. More broadly, this work reinforces the value of continued instrument development at XFELs: by enabling high-quality total scattering and PDF measurements in increasingly realistic sample environments, these advances open the door to new scientific breakthroughs in chemistry, materials science, and ultrafast structural dynamics.

## Acknowledgements

We acknowledge European XFEL in Schenefeld, Germany, for the provision of XFEL beam time at Scientific Instrument MID (Materials Imaging and Dynamics) under proposal number 8776 and would like to thank the staff of European XFEL and DESY for their assistance. The authors thank the HIBEF user consortium for providing instrumentation and staff during this experiment. AFS acknowledges MAX IV Laboratory for time on Beamline DanMAX (proposal No. 20230435) and would like to thank Mads R. V. Jørgensen and Frederik H. Gjørup for their assistance in collecting the synchrotron data on ferrihydrite. Research conducted at MAX IV is supported by the Swedish Research Council under contract 2018–07152, the Swedish Governmental Agency for Innovation Systems under contract 2018–04969 and Formas under contract 2019–02496. DanMAX is funded by the NUFI Grant Number 4059-00009B. FB and AG would like to acknowledge the European Synchrotron Radiation Facility (Grenoble, France) for providing beamtime to collect reference synchrotron data on lead halide perovskites under proposal number CH-7356 (DOI: 10.15151/ESRF-ES-2041039507) and the ID22 High-Resolution Powder Diffraction Beamline staff for their assistance and support.

## Data Availability

Data recorded for the experiment at the European XFEL are available at https://doi.org/10.22003/XFEL.EU-DATA-008776-00.

## Funding Information

AFS and KMØJ acknowledge the Novo Nordisk Foundation (0085640). We thank the Danish Agency for Science, Technology, and Innovation for funding the instrument centre DanScatt. LJS, ABB, VAK, and BBI acknowledge the Novo Nordisk Foundation (NNF230C0085064) and the Carlsberg Foundation (CF24-1527). EAH acknowledges the UK Hub for the Physical Sciences on XFELs (HPSX) for part-funding her DPhil Studentship. ESB was supported by the European Union's Horizon Europe research and innovation programme under grant agreement No. 101185375.

## References

1 D. A. Keen, *Crystallogr. Rev.*, 2020, **26**, 143–201.
2 T. Egami and S. J. L. Billinge, *Underneath the Bragg Peaks*, Elsevier 2012.
3 M. W. Terban and S. J. L. Billinge, *Chem. Rev.*, 2022, **122**, 1208–1272.
4 D. A. Keen, *J. Appl. Cryst.*, 2001, **34**, 172–177.
5 N. P. L. Magnard, D. R. Sørensen, I. Kantor, K. M. Ø. Jensen and M. R. V. Jørgensen, *J. Appl. Cryst.*, 2023, **56**, 825–833.
6 K. H. Kim, J. G. Kim, S. Nozawa, T. Sato, K. Y. Oang, T. W. Kim, H. Ki, J. Jo, S. Park, C. Song, T. Sato, K. Ogawa, T. Togashi, K. Tono, M. Yabashi, T. Ishikawa, J. Kim, R. Ryoo, J. Kim, H. Ihee and S. Adachi, *Nature*, 2015, **518**, 385–389.
7 A. Rousse, C. Rischel, S. Fourmaux, I. Uschmann, S. Sebban, G. Grillon, P. Balcou, E. Förster, J. P. Geindre, P. Audebert, J. C. Gauthier and D. Hulin, *Nature*, 2001, **410**, 65–68.
8 P. Beaud, A. Caviezel, S. O. Mariager, L. Rettig, G. Ingold, C. Dornes, S.-W. Huang, J. A. Johnson, M. Radovic, T. Huber, T. Kubacka, A. Ferrer, H. T. Lemke, M. Chollet, D. Zhu, J. M. Glownia, M. Sikorski, A. Robert, H. Wadati, M. Nakamura, M. Kawasaki, Y. Tokura, S. L. Johnson and U. Staub, *Nat. Mater.*, 2014, **13**, 923–927.
9 B. Guzelturk, T. Winkler, T. W. J. Van De Goor, M. D. Smith, S. A. Bourelle, S. Feldmann, M. Trigo, S. W. Teitelbaum, H.-G. Steinrück, G. A. De La Pena, R. Alonso-Mori, D. Zhu, T. Sato, H. I. Karunadasa, M. F. Toney, F. Deschler and A. M. Lindenberg, *Nat. Mater.*, 2021, **20**, 618–623.
10 J. Griffiths, A. F. Suzana, L. Wu, S. D. Marks, V. Esposito, S. Boutet, P. G. Evans, J. F. Mitchell, M. P. M. Dean, D. A. Keen, I. Robinson, S. J. L. Billinge and E. S. Bozin, *Nat. Mater.*, 2024, **23**, 1041–1047.
11 I. Georgescu, *Nat. Rev. Phys.*, 2020, **2**, 345–345.
12 C. Crépisson, M. Fitzgerald, D. Peake, P. G. Heighway, T. Stevens, A. Descamps, D. McGonegle, A. Amouretti, K. K. Alaa El-Din, M. Andrzejewski, S. Azadi, E. Brambrink, C. Camarda, D. A. Chin, S. Di Dio Cafiso, A. Coutinho Dutra, H. Höppner, K. Yamamoto, P. S. Karamched, Z. Konôpková, M. Nakatsutsumi, N. Ozaki, D. N. Polsin, J.-P. Schwinkendorf, G. Shoulga, C. Strohm, M. Tang, H. Taylor, M. Toncian, Y. Wang, J. Yao, G. Gregori, J. S. Wark, K. Appel, M. Harmand and S. M. Vinko, *Nat. Commun.*, 2026, **17**, 8527.
13 D. Kraus, J. Rips, M. Schörner, M. G. Stevenson, J. Vorberger, D. Ranjan, J. Lütgert, B. Heuser, J. H. Eggert, H.-P. Liermann, I. I. Oleynik, S. Pandolfi, R. Redmer, A. Sollier, C. Strohm, T. J. Volz, B. Albertazzi, S. J. Ali, L. Antonelli, C. Bähtz, O. B. Ball, S. Banerjee, A. B. Belonoshko, C. A. Bolme, V. Bouffetier, R. Briggs, K. Buakor, T. Butcher, V. Cerantola, J. Chantel, A. L. Coleman, J. Collier, G. W. Collins, A. J. Comley, T. E. Cowan, G. Cristoforetti, H. Cynn, A. Descamps, A. Di Cicco, S. Di Dio Cafiso, F. Dorchies, M. J. Duff, A. Dwivedi, C. Edwards, D. Errandonea, S. Galitskiy, E. Galtier, H. Ginestet, L. Gizzi, A. Gleason, S. Göde, J. M. Gonzalez, M. G. Gorman, M. Harmand, N. J. Hartley, P. G. Heighway, C. Hernandez-Gomez, A. Higginbotham, H. Höppner, R. J. Husband, T. M. Hutchinson, H. Hwang, D. A. Keen, J. Kim, P. Koester, Z. Konôpková, A. Krygier, L. Labate, A. Laso Garcia, A. E. Lazicki, Y. Lee, P. Mason, M. Masruri, B. Massani, E. E. McBride, J. D. McHardy, D. McGonegle, C. McGuire, R. S. McWilliams, S. Merkel, G. Morard, B. Nagler, M. Nakatsutsumi, K. Nguyen-Cong, A.-M. Norton, N. Ozaki, C. Otzen, D. J. Peake, A. Pelka, K. A. Pereira, J. P. Phillips, C. Prescher, T. R. Preston, L. Randolph, A. Ravasio, D. Santamaria-Perez, D. J. Savage, M. Schölmerich, J.-P. Schwinkendorf, S. Singh, J. Smith, R. F. Smith, J. Spear, C. Spindloe, T.-A. Suer, M. Tang, M. Toncian, T. Toncian, S. J. Tracy, A. Trapananti, C. E. Vennari, T. Vinci, M. Tyldesley, S. C. Vogel, J. P. S. Walsh, J. S. Wark, J. T. Willman, L. Wollenweber, U. Zastrau, E. Brambrink, K. Appel and M. I. McMahon, *Nature*, 2025, **642**, 351–355.

14 M. G. Gorman, D. McGonegle, R. F. Smith, S. Singh, T. Jenkins, R. S. McWilliams, B. Albertazzi, S. J. Ali, L. Antonelli, M. R. Armstrong, C. Baehtz, O. B. Ball, S. Banerjee, A. B. Belonoshko, A. Benuzzi-Mounaix, C. A. Bolme, V. Bouffetier, R. Briggs, K. Buakor, T. Butcher, S. Di Dio Cafiso, V. Cerantola, J. Chantel, A. Di Cicco, S. Clarke, A. L. Coleman, J. Collier, G. W. Collins, A. J. Comley, F. Coppari, T. E. Cowan, G. Cristoforetti, H. Cynn, A. Descamps, F. Dorchies, M. J. Duff, A. Dwivedi, C. Edwards, J. H. Eggert, D. Errandonea, G. Fiquet, E. Galtier, A. Laso Garcia, H. Ginestet, L. Gizzi, A. Gleason, S. Goede, J. M. Gonzalez, M. Harmand, N. J. Hartley, P. G. Heighway, C. Hernandez-Gomez, A. Higginbotham, H. Höppner, R. J. Husband, T. M. Hutchinson, H. Hwang, A. E. Lazicki, D. A. Keen, J. Kim, P. Koester, Z. Konopkova, D. Kraus, A. Krygier, L. Labate, Y. Lee, H.-P. Liermann, P. Mason, M. Masruri, B. Massani, E. E. McBride, C. McGuire, J. D. McHardy, S. Merkel, G. Morard, B. Nagler, M. Nakatsutsumi, K. Nguyen-Cong, A.-M. Norton, I. I. Oleynik, C. Otzen, N. Ozaki, S. Pandolfi, D. J. Peake, A. Pelka, K. A. Pereira, J. P. Phillips, C. Prescher, T. R. Preston, L. Randolph, D. Ranjan, A. Ravasio, R. Redmer, J. Rips, D. Santamaria-Perez, D. J. Savage, M. Schoelmerich, J.-P. Schwinkendorf, J. Smith, A. Sollier, J. Spear, C. Spindloe, M. Stevenson, C. Strohm, T.-A. Suer, M. Tang, M. Toncian, T. Toncian, S. J. Tracy, A. Trapananti, T. Tschentscher, M. Tyldesley, C. E. Vennari, T. Vinci, S. C. Vogel, T. J. Volz, J. Vorberger, J. P. S. Walsh, J. S. Wark, J. T. Willman, L. Wollenweber, U. Zastrau, E. Brambrink, K. Appel and M. I. McMahon, *J. Appl. Phys.*, 2024, **135**, 165902.

15 L. J. Støckler, R. S. Christensen, M. Kløve, A. D. Bertelsen, A. B. Borup, L. Krause, S. Takahashi, T. Fujita, H. Kasai, I. Inoue, E. Nishibori and B. B. Iversen, *ChemPhysChem*, 2023, **24**, e202300407.

16 A. F. Sapnik, P. A. Chater, D. S. Keeble, J. S. O. Evans, F. Bertolotti, A. Guagliardi, L. J. Støckler, E. A. Harbourne, A. B. Borup, R. S. Silberg, A. Descamps, C. Prescher, B. D. Klee, A. Phelipeau, I. Ullah, K. G. Medina, T. A. Bird, V. Kaznelson, W. Lynn, A. L. Goodwin, B. B. Iversen, C. Crepisson, E. S. Bozin, K. M. Ø. Jensen, E. E. McBride, R. B. Neder, I. Robinson, J. S. Wark, M. Andrzejewski, U. Boesenberg, E. Brambrink, C. Camarda, V. Cerantola, S. Goede, H. Höppner, O. S. Humphries, Z. Konopkova, N. Kujala, T. Michelat, M. Nakatsutsumi, A. Pelka, T. R. Preston, L. Randolph, M. Roeper, A. Schmidt, C. Strohm, M. Tang, P. Talkovski, U. Zastrau, K. Appel and D. A. Keen, *IUCrJ*, 2025, **12**, 531–547.

17 H. Weise and W. Decking, *Proceedings of the 38th Int. Free Electron Laser Conf.*, 2018, MOC03, 9–13.

18 W. Decking, S. Abeghyan, P. Abramian, A. Abramsky, A. Aguirre, C. Albrecht, P. Alou, M. Altarelli, P. Altmann, K. Amyan, V. Anashin, E. Apostolov, K. Appel, D. Auguste, V. Ayvazyan, S. Baark, F. Babies, N. Baboi, P. Bak, V. Balandin, R. Baldinger, B. Baranasic, S. Barbanotti, O. Belikov, V. Belokurov, L. Belova, V. Belyakov, S. Berry, M. Bertucci, B. Beutner, A. Block, M. Blöcher, T. Böckmann, C. Bohm, M. Böhnert, V. Bondar, E. Bondarchuk, M. Bonezzi, P. Borowiec, C. Bösch, U. Bösenberg, A. Bosotti, R. Böspflug, M. Bousonville, E. Boyd, Y. Bozhko, A. Brand, J. Branlard, S. Briechle, F. Brinker, S. Brinker, R. Brinkmann, S. Brockhauser, O. Brovko, H. Brück, A. Brüdgam, L. Butkowski, T. Büttner, J. Calero, E. Castro-Carballo, G. Cattalanotto, J. Charrier, J. Chen, A. Cherepenko, V. Cheskidov, M. Chiodini, A. Chong, S. Choroba, M. Chorowski, D. Churanov, W. Cichalewski, M. Clausen, W. Clement, C. Cloué, J. A. Cobos, N. Coppola, S. Cunis, K. Czuba, M. Czwalinna, B. D'Almagne, J. Dammann, H. Danared, A. de Zubiaurre Wagner, A. Delfs, T. Delfs, F. Dietrich, T. Dietrich, M. Dohlus, M. Dommach, A. Donat, X. Dong, N. Doynikov, M. Dressel, M. Duda, P. Duda, H. Eckoldt, W. Ehsan, J. Eidam, F. Eints, C. Engling, U. Englisch, A. Ermakov, K. Escherich, J. Eschke, E. Saldin, M. Faesing, A. Fallou, M. Felber, M. Fenner, B. Fernandes, J. M. Fernández, S. Feuker, K. Filippakopoulos, K. Floettmann, V. Fogel, M. Fontaine, A. Francés, I. F. Martin, W. Freund, T. Freyermuth, M. Friedland, L. Fröhlich, M. Fusetti, J. Fydrych, A. Gallas, O. García, L. Garcia-Tabares, G.

Geloni, N. Gerasimova, C. Gerth, P. Geßler, V. Gharibyan, M. Gloor, J. Głowinkowski, A. Goessel, Z. Gołębiewski, N. Golubeva, W. Grabowski, W. Graeff, A. Grebentsov, M. Grecki, T. Grevsmuehl, M. Gross, U. Grosse-Wortmann, J. Grünert, S. Grunewald, P. Grzegory, G. Feng, H. Guler, G. Gusev, J. L. Gutierrez, L. Hagge, M. Hamberg, R. Hanneken, E. Harms, I. Hartl, A. Hauberg, S. Hauf, J. Hauschildt, J. Hauser, J. Havlicek, A. Hedqvist, N. Heidbrook, F. Hellberg, D. Henning, O. Hensler, T. Hermann, A. Hidvégi, M. Hierholzer, H. Hintz, F. Hoffmann, M. Hoffmann, M. Hoffmann, Y. Holler, M. Hüning, A. Ignatenko, M. Ilchen, A. Iluk, J. Iversen, J. Iversen, M. Izquierdo, L. Jachmann, N. Jardon, U. Jastrow, K. Jensch, J. Jensen, M. Jeżabek, M. Jidda, H. Jin, N. Johansson, R. Jonas, W. Kaabi, D. Kaefer, R. Kammering, H. Kapitza, S. Karabekyan, S. Karstensen, K. Kasprzak, V. Katalev, D. Keese, B. Keil, M. Kholopov, M. Killenberger, B. Kitaev, Y. Klimchenko, R. Klos, L. Knebel, A. Koch, M. Koepke, S. Köhler, W. Köhler, N. Kohlstrunk, Z. Konopkova, A. Konstantinov, W. Kook, W. Koprek, M. Körfer, O. Korth, A. Kosarev, K. Kosiński, D. Kostin, Y. Kot, A. Kotarba, T. Kozak, V. Kozak, R. Kramert, M. Krasilnikov, A. Krasnov, B. Krause, L. Kravchuk, O. Krebs, R. Kretschmer, J. Kreutzkamp, O. Kröplin, K. Krzysik, G. Kube, H. Kuehn, N. Kujala, V. Kulikov, V. Kuzminych, D. La Civita, M. Lacroix, T. Lamb, A. Lancetov, M. Larsson, D. Le Pinvidic, S. Lederer, T. Lensch, D. Lenz, A. Leuschner, F. Levenhagen, Y. Li, J. Liebing, L. Lilje, T. Limberg, D. Lipka, B. List, J. Liu, S. Liu, B. Lorbeer, J. Lorkiewicz, H. H. Lu, F. Ludwig, K. Machau, W. Maciocha, C. Madec, C. Magueur, C. Maiano, I. Maksimova, K. Malcher, T. Maltezopoulos, E. Mamoshkina, B. Manschwetus, F. Marcellini, G. Marinkovic, T. Martinez, H. Martirosyan, W. Maschmann, M. Maslov, A. Matheisen, U. Mavric, J. Meißner, K. Meissner, M. Messerschmidt, N. Meyners, G. Michalski, P. Michelato, N. Mildner, M. Moe, F. Moglia, C. Mohr, S. Mohr, W. Möller, M. Mommerz, L. Monaco, C. Montiel, M. Moretti, I. Morozov, P. Morozov, D. Mross, J. Mueller, C. Müller, J. Müller, K. Müller, J. Munilla, A. Münnich, V. Muratov, O. Napoly, B. Näser, N. Nefedov, R. Neumann, R. Neumann, N. Ngada, D. Noelle, F. Obier, I. Okunev, J. A. Oliver, M. Omet, A. Oppelt, A. Ottmar, M. Oublaid, C. Pagani, R. Paparella, V. Paramonov, C. Peitzmann, J. Penning, A. Perus, F. Peters, B. Petersen, A. Petrov, I. Petrov, S. Pfeiffer, J. Pflüger, S. Philipp, Y. Pienaud, P. Pierini, S. Pivovarov, M. Planas, E. Pławski, M. Pohl, J. Polinski, V. Popov, S. Prat, J. Prenting, G. Priebe, H. Pryschelski, K. Przygoda, E. Pyata, B. Racky, A. Rathjen, W. Ratuschni, S. Regnaud-Campderros, K. Rehlich, D. Reschke, C. Robson, J. Roever, M. Roggli, J. Rothenburg, E. Rusiński, R. Rybaniec, H. Sahling, M. Salmani, L. Samoylova, D. Sanzone, F. Saretzki, O. Sawlanski, J. Schaffran, H. Schlarb, M. Schlösser, V. Schlott, C. Schmidt, F. Schmidt-Foehre, M. Schmitz, M. Schmökel, T. Schnautz, E. Schneidmiller, M. Scholz, B. Schöneburg, J. Schultze, C. Schulz, A. Schwarz, J. Sekutowicz, D. Sellmann, E. Semenov, S. Serkez, D. Sertore, N. Shehzad, P. Shemarykin, L. Shi, M. Sienkiewicz, D. Sikora, M. Sikorski, A. Silenzi, C. Simon, W. Singer, X. Singer, H. Sinn, K. Sinram, N. Skvorodnev, P. Smirnow, T. Sommer, A. Sorokin, M. Stadler, M. Steckel, B. Steffen, N. Steinhau-Kühl, F. Stephan, M. Stodulski, M. Stolper, A. Sulimov, R. Susen, J. Świerblewski, C. Sydlo, E. Syresin, V. Sytchev, J. Szuba, N. Tesch, J. Thie, A. Thiebault, K. Tiedtke, D. Tischhauser, J. Tolkiehn, S. Tomin, F. Tonisch, F. Toral, I. Torbin, A. Trapp, D. Treyer, G. Trowitzsch, T. Trublet, T. Tschentscher, F. Ullrich, M. Vannoni, P. Varela, G. Varghese, G. Vashchenko, M. Vasic, C. Vazquez-Velez, A. Verguet, S. Vilcins-Czvitkovits, R. Villanueva, B. Visentin, M. Viti, E. Vogel, E. Volobuev, R. Wagner, N. Walker, T. Wamsat, H. Weddig, G. Weichert, H. Weise, R. Wenndorf, M. Werner, R. Wichmann, C. Wiebers, M. Wiencek, T. Wilksen, I. Will, L. Winkelmann, M. Winkowski, K. Wittenburg, A. Witzig, P. Wlk, T. Wohlenberg, M. Wojciechowski, F. Wolff-Fabris, G. Wrochna, K. Wrona, M. Yakopov, B. Yang, F. Yang, M. Yurkov, I. Zagorodnov, P. Zalden, A. Zavadtsev, D. Zavadtsev, A. Zhirnov, A. Zhukov, V. Ziemann, A. Zolotov, N. Zolotukhina, F. Zummack and D. Zybin, *Nat. Photonics*, 2020, **14**, 391–397.

19 A. Madsen, J. Hallmann, G. Ansaldi, T. Roth, W. Lu, C. Kim, U. Boesenberg, A. Zozulya, J. Möller, R. Shayduk, M. Scholz, A. Bartmann, A. Schmidt, I. Lobato, K. Sukharnikov, M. Reiser, K. Kazarian and I. Petrov, *J. Synchrotron Radiat.*, 2021, **28**, 637–649.
20 N. Burns, A. Rahemtulla, S. Annett, B. Moreno and S. Kycia, *J. Appl. Cryst.*, 2023, **56**, 510–518.
21 F. A. Lima, F. Otte, M. Vakili, F. Ardana-Lamas, M. Biednov, F. Dall'Antonia, P. Frankenberger, W. Gawelda, L. Gelisio, H. Han, X. Huang, Y. Jiang, M. Kloos, T. Kluyver, M. Knoll, K. Kubicek, I. J. Bermudez Macias, J. Schulz, O. Turkot, Y. Uemura, J. Valerio, H. Wang, H. Yousef, P. Zalden, D. Khakhulin, C. Bressler and C. Milne, *J. Synchrotron Radiat.*, 2023, **30**, 1168–1182.
22 A. Gierke, C. Goy, A. Schmidt, J. Möller, M. Walther, R. P. C. Bauer, J.-E. Pudell, W. Jo, J. Valerio, A. Wrona, J. Schulz, A. Madsen and F. Lehmkühler, *J. Phys.: Conf. Ser.*, 2025, **3010**, 012151.
23 M. Pergament, G. Palmer, M. Kellert, K. Kruse, J. Wang, L. Wissmann, U. Wegner, M. Emons, D. Kane, G. Priebe, S. Venkatesan, T. Jezynski, F. Pallas and M. J. Lederer, *Opt. Express*, 2016, **24**, 29349.
24 N. Yazdani, M. I. Bodnarchuk, F. Bertolotti, N. Masciocchi, I. Fureraj, B. Guzelturk, B. L. Cotts, M. Zajac, G. Rainò, M. Jansen, S. C. Boehme, M. Yarema, M.-F. Lin, M. Kozina, A. Reid, X. Shen, S. Weathersby, X. Wang, E. Vauthey, A. Guagliardi, M. V. Kovalenko, V. Wood and A. M. Lindenberg, *Nat. Phys.*, 2024, **20**, 47–53.
25 G. Ashiotis, A. Deschildre, Z. Nawaz, J. P. Wright, D. Karkoulis, F. E. Picca and J. Kieffer, *J. Appl. Cryst.*, 2015, **48**, 510–519.
26 A. A. Coelho, *J. Appl. Cryst.*, 2018, **51**, 210–218.
27 A. Cervellino, R. Frison, F. Bertolotti and A. Guagliardi, *J. Appl. Cryst.*, 2015, **48**, 2026–2032.
28 A. Cervellino, C. Giannini and A. Guagliardi, *J. Comput. Chem.*, 2006, **27**, 995–1008.
29 P. Juhás, T. Davis, C. L. Farrow and S. J. L. Billinge, *J. Appl. Cryst.*, 2013, **46**, 560–566.
30 C. L. Farrow, P. Juhas, J. W. Liu, D. Bryndin, E. S. Božin, J. Bloch, T. Proffen and S. J. L. Billinge, *J. Phys.: Condens. Matter*, 2007, **19**, 335219.
31 P. Juhás, C. Farrow, X. Yang, K. Knox and S. Billinge, *Acta Cryst. A*, 2015, **71**, 562–568.
32 A. K. Soper, *Chem. Phys.*, 1996, **202**, 295–306.
33 A. K. Soper, *Mol. Phys.*, 2001, **99**, 1503–1516.
34 A. K. Soper, *Phys. Rev. B*, 2005, **72**, 104204.
35 A. K. Soper, *GudrunN and GudrunX: Programs for Correcting Raw Neutron and X-ray Diffraction Data to Differential Scattering Cross Section*, Technical report RAL-TR-2011-013, 2011.
36 D. A. Keen, *J. Appl. Cryst.*, 2001, **34**, 172–177.
37 S. J. Smith, K. Page, H. Kim, B. J. Campbell, J. Boerio-Goates and B. F. Woodfield, *Inorg. Chem.*, 2012, **51**, 6421–6424.
38 Q. A. Akkerman, T. P. T. Nguyen, S. C. Boehme, F. Montanarella, D. N. Dirin, P. Wechsler, F. Beiglböck, G. Rainò, R. Erni, C. Katan, J. Even and M. V. Kovalenko, *Science*, 2022, **377**, 1406–1412.
39 M. Yashima and T. Takizawa, *J. Phys. Chem. C*, 2010, **114**, 2385–2392.
40 J. M. Zuo, P. Blaha and K. Schwarz, *J. Phys.: Condens. Matter*, 1997, **9**, 7541–7561.
41 J. Beyer, N. Roth and B. Brummerstedt Iversen, *Acta Crystallogr. A Found. Adv.*, 2022, **78**, 10–20.
42 F. M. Michel, L. Ehm, S. M. Antao, P. L. Lee, P. J. Chupas, G. Liu, D. R. Strongin, M. A. A. Schoonen, B. L. Phillips and J. B. Parise, *Science*, 2007, **316**, 1726–1729.
43 N. P. Funnell, M. F. Fulford, S. Inoué, K. Kletetschka, F. M. Michel and A. L. Goodwin, *Commun. Chem.*, 2020, **3**, 22.

44 U. Schwertmann and R. M. Cornell, Eds, *Iron Oxides in the Laboratory*, Wiley-VCH Verlag GmbH, Weinheim, Germany, 2000.

45 A. Nimmrich, M. R. Panman, O. Berntsson, E. Biasin, S. Niebling, J. Petersson, M. Hoernke, A. Björling, E. Gustavsson, T. B. Van Driel, A. O. Dohn, M. Laursen, D. B. Zederkof, K. Tono, T. Katayama, S. Owada, M. M. Nielsen, J. Davidsson, J. Uhlig, J. S. Hub, K. Haldrup and S. Westenhoff, *J. Am. Chem. Soc.*, 2023, **145**, 15754–15765.

46 R. P. C. Bauer, A. Gierke, C. Goy, J.-E. Pudell, W. Jo, J. Valerio, L. Adriano, J. Möller, J. Wrigley, U. Boesenberg, A. Rodriguez-Fernandez, R. Shayduk, M. Youssef, N. N. Striker, N. C. Giesselmann, S. C. Hövelmann, W. Roseker, K. Ayyer, M. Paulus, C. Sternemann, R. E. Grisenti, A. Madsen, J. Schulz and F. Lehmkühler, *J. Phys.: Conf. Ser.*, 2025, **3010**, 012156.

47 A. K. Soper, *ISRN Phys. Chem.*, 2013, **2013**, 1–67.

48 J. Yang, R. Dettori, J. P. F. Nunes, N. H. List, E. Biasin, M. Centurion, Z. Chen, A. A. Cordones, D. P. Deponte, T. F. Heinz, M. E. Kozina, K. Ledbetter, M.-F. Lin, A. M. Lindenberg, M. Mo, A. Nilsson, X. Shen, T. J. A. Wolf, D. Donadio, K. J. Gaffney, T. J. Martinez and X. Wang, *Nature*, 2021, **596**, 531–535.

49 F. Bertolotti, L. Protesescu, M. V. Kovalenko, S. Yakunin, A. Cervellino, S. J. L. Billinge, M. W. Terban, J. S. Pedersen, N. Masciocchi and A. Guagliardi, *ACS Nano*, 2017, **11**, 3819–3831.

50 F. Bertolotti, N. Dengo, A. Cervellino, M. I. Bodnarchuk, C. Bernasconi, I. Cherniukh, Y. Berezovska, S. C. Boehme, M. V. Kovalenko, N. Masciocchi and A. Guagliardi, *Small Structures*, 2024, **5**, 2300264.

51 H. Ihee, *Acc. Chem. Res.*, 2009, **42**, 356–366.

52 M. R. Panman, E. Biasin, O. Berntsson, M. Hermann, S. Niebling, A. J. Hughes, J. Kübel, K. Atkovska, E. Gustavsson, A. Nimmrich, A. O. Dohn, M. Laursen, D. B. Zederkof, A. Honarfar, K. Tono, T. Katayama, S. Owada, T. B. Van Driel, K. Kjaer, M. M. Nielsen, J. Davidsson, J. Uhlig, K. Haldrup, J. S. Hub and S. Westenhoff, *Phys. Rev. Lett.*, 2020, **125**, 226001.

53 R. P. Kurta, T. B. Van Driel, A. O. Dohn, T. B. Berberich, S. Nelson, I. A. Zaluzhnyy, N. Mukharamova, D. Lapkin, D. B. Zederkof, M. Seaberg, K. S. Pedersen, K. S. Kjær, G. I. Rippy, E. Biasin, K. B. Møller, L. Gelisio, K. Haldrup, I. A. Vartanyants and M. M. Nielsen, *Phys. Chem. Chem. Phys.*, 2023, **25**, 23417–23434.

54 V. Markmann, J. Pan, B. L. Hansen, M. L. Haubro, A. Nimmrich, P. Lenzen, M. Levantino, T. Katayama, S. Adachi, S. Gorski-Bilke, F. Temps, A. O. Dohn, K. B. Møller, M. M. Nielsen and K. Haldrup, *Chem. Sci.*, 2024, **15**, 11391–11401.

55 B. L. Hansen, V. Markmann, M. Pápai, P. Lenzen, M. L. Haubro, A. Mikeházi, Z. Németh, A. Vancza, M. Levantino, S. Zerdane, D. B. Zederkof, D. Khakhulin, A. O. Dohn, M. M. Nielsen, K. Haldrup and G. Vankó, *J. Chem. Phys.*, 2025, **162**, 124308.

56 A. K. Soper, *J. Phys.: Condens. Matter*, 2007, **19**, 335206.

57 D. T. Bowron and S. D. Moreno, *J. Phys.: Condens. Matter*, 2013, **25**, 454213.

58 A. Bernasconi, M. Dapiaggi, A. Pavese, D. T. Bowron and S. Imberti, *J. Phys. Chem. B*, 2012, **116**, 13114–13123.

59 A. K. Soper, *Mol. Phys.*, 2009, **107**, 1667–1684.

60 J. Schulz, J. Bielecki, R. B. Doak, K. Dörner, R. Graceffa, R. L. Shoeman, M. Sikorski, P. Thute, D. Westphal and A. P. Mancuso, *J. Synchrotron Radiat.*, 2019, **26**, 339–345.

61 P. E. Konold, A. Ansari, D. Westphal, P. Adams, R. Letrun, J. Valerio, J. Bielecki, R. C. Alvarez, S. Bódizs, K. Dörner, G. Ducrocq, L. Grunewald, K. Kharitonov, C. Kim, M. Kloos, A. Králová, T. Larkiala, M. Maihöfer, D. Manatou, R. Mehlig, D. V. M. Melo, P. Meszaros, L. Monrroy, T. Sato, R. Schubert, P. Smyth, E. Sobolev, F. Trost, R. De Wijn, L. Worbs, T. V. Yenupuri, T. You, J. Schulz, R. Bean, K. Ayyer, M. A. D. Martin, H. Kirst, H. Grubmüller, S. Westenhoff, R. A. Kirian and F. R. N. C. Maia, *IUCrJ*, 2026, **13**, 659–667.

62 C. Antolini, V. Sosa Alfaro, M. Reinhard, G. Chatterjee, R. Ribson, D. Sokaras, L. Gee, T. Sato, P. L. Kramer, S. L. Raj, B. Hayes, P. Schleissner, A. T. Garcia-Esparza, J. Lim, J. T. Babicz, A. H. Follmer, S. Nelson, M. Chollet, R. Alonso-Mori and T. B. Van Driel, *Molecules*, 2024, **29**, 2323.

# Ultrafast Liquid Jet X-ray Total Scattering

## SUPPLEMENTARY INFORMATION

Adam F. Sapnik,[a]* Philip A. Chater,[b] Dean S. Keeble,[b] Tobias A. Bird,[b] John S. O. Evans,[c] Federica Bertolotti,[d] Antonietta Guagliardi,[e] Norberto Masciocchi,[d] Daniel T. Bowron,[f] Lise J. Støckler,[g] Elodie A. Harbourne,[h] Anders B. Borup,[g] Rebecca S. Silberg,[a] Viktoria A. Kaznelson,[g] Stefanie Punke,[a] Mariia Svyrydenko,[i,j] Maryna I. Bodnarchuk,[i,j] Dmitry N. Dirin,[i,j] Simon C. Boehme,[j,i] Maksym V. Kovalenko,[i,j] Bo B. Iversen,[g] Emil S. Bozin,[k] Kirsten M. Ø. Jensen,[a] Gabriele Ansaldi,[l] Karen Appel,[l] Alexander Gierke,[l,m] Jörg Hallmann,[l] Aliaksandr Leonau,[l] Anders Madsen,[l] Johannes Möller,[l] Jan-Etienne Pudell,[l] Angel Rodriguez-Fernandez,[l] Andreas Schmidt,[l] Roman Shayduk,[l] Cornelius Strohm,[m] Joana Valerio,[l] Ulrike Boesenberg,[l] and David A. Keen.[f]*

Email: afs@chem.ku.dk & david.keen@stfc.ac.uk

a. Department of Chemistry, University of Copenhagen, Universitetsparken 5, 2100 Copenhagen Ø, Denmark.
b. Diamond Light Source Ltd, Diamond House, Harwell Science & Innovation Campus, Didcot, Oxfordshire, UK.
c. Department of Chemistry, University Science Site, Durham University, South Road, Durham, DH1 3LE, UK.
d. Dipartimento di Scienze e Alta Tecnologia, University of Insubria, 22100, Como, Italy; To.Sca.Lab, University of Insubria, 22100, Como, Italy.
e. Instituo di Cristallografia, Consiglio Nazionale delle Ricerche (CNR), 22100, Como, Italy; To.Sca.Lab, University of Insubria, 22100, Como, Italy.
f. ISIS Facility, Rutherford Appleton Laboratory, Harwell Campus, Didcot, Oxfordshire OX11 0QX, UK.
g. Center for Integrated Materials Research, Department of Chemistry and iNANO, Aarhus University, Langelandsgade 140, 8000 Aarhus C, Denmark.
h. Department of Chemistry, Inorganic Chemistry Laboratory, University of Oxford, South Parks Road, Oxford OX1 3QR, UK.
i. EMPA−Swiss Federal Laboratories for Materials Science and Technology, Dübendorf, Switzerland.
j. Institute of Inorganic Chemistry, Department of Chemistry and Applied Biosciences, ETH Zürich, Zürich, Switzerland.
k. Center for Solid State Physics and New Materials, Institute of Physics Belgrade, University of Belgrade, Pregrevica 118, 11080 Belgrade, Serbia.
l. European X-Ray Free-Electron Laser Facility, Holzkoppel 4, 22869 Schenefeld, Germany.
m. Deutsches Elektronen-Synchrotron DESY, Notkestrasse 85, 22607 Hamburg, Germany.

# Supplementary Information

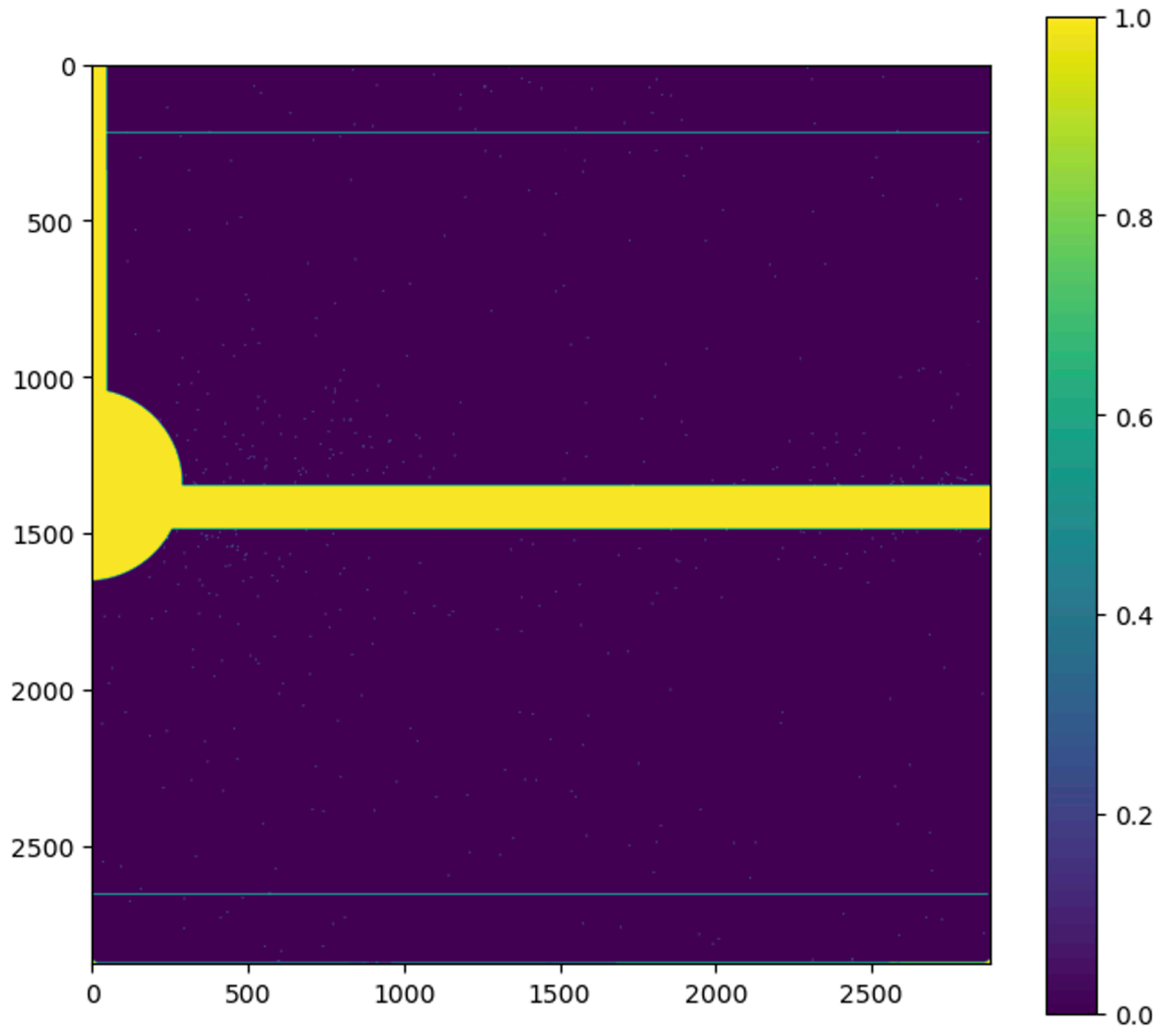

**Supplementary Figure 1** Combined detector mask constructed by combining a manually defined masked detector region (including that which was directly behind the liquid jet), the manufacturer bad-pixel mask and pixels with invalid gain or solid-angle correction values.

**Supplementary Table 1** Width of the liquid jet as a function of flow rate, using a He pressure of 200 PSI. Measurements marked with *a* or *b* were performed at 245 and 220 PSI He, respectively.

| Flow rate (mL min$^{-1}$) | $H_2O$ jet (µm) | Ethanol jet (µm) | 0.5 M Si in $H_2O$ jet (µm) |
|---|---|---|---|
| 0.3 | 52[a] | 55 | 53[a] |
| 0.4 | 64[b] | 61 | 64[b] |
| 0.5 | 72 | 66 | 73 |
| 0.6 | 77 | 70 | 78 |
| 0.7 | 80 | 74 | 80 |
| 0.8 | 83 | 79 | 84 |
| 0.9 | 85 | 83 | 88 |
| 1.0 | 89 | 85 | 90 |

**Supplementary Table 2** Sample details and processing parameters used for obtaining the PDF.

| | $CeO_2$ | Si | Ferrihydrite | Water | Si | Keggin |
|---|---|---|---|---|---|---|
| **State** | Powder | Powder | Powder | Liquid | Suspension | Solution |
| **Solvent** | - | - | - | - | Water | Water |
| **Delivery** | Capillary | Capillary | Capillary | LJ | LJ | LJ |
| **Composition** | $CeO_2$ | Si | $Fe_2O_{3.5}H$ | $H_2O$ | Si | $H_2W_{12}O_{40}$ |
| **Background** | Capillary | Capillary | Capillary | Air | Air | Water & air |
| $Q_{max,inst}$ (Å$^{-1}$) | 16 | 16 | 16 | 16 | 14 | 14.2 |
| $Q_{max}$ (Å$^{-1}$) | 16 | 16 | 16 | 12 | 12 | 14.1 |
| $Q_{min}$ (Å$^{-1}$) | 1 | 1 | 1.6 | 1 | 1.5 | 1 |
| $r_{poly}$ | 0.9 | 0.9 | 1.6 | 1.4 | 1.4 | 1.5 |

**Supplementary Table 3** Reference potential parameters used in the EPSR simulation. (1) Wk: tungsten atoms of the Keggin, (2) Ok: oxygen atoms of the Keggin that bridge between tungsten atoms, (3) Okk: oxygen atoms of the Keggin that are oriented outwards from the cluster surface, (4) Oc: oxygen atoms at the core of the Keggin, (5) Nam: nitrogen atoms of the ammonium ions in solution, (6) Ham: hydrogen atoms of the ammonium ions in solution, (7) Ow: water oxygen atoms, and (8) Hw: water hydrogen atoms.

| Atom type | Atomic mass (u) | Lennard-Jones energy (ε) (kJ/mol) | Lennard-Jones atom size (σ) (Å) | Coulomb charge (q) (e) |
|---|---|---|---|---|
| Wk | 183.95 | 0.65 | 3.400 | 0.0000 |
| Ok | 16.00 | 0.65 | 3.070 | 0.0000 |
| Okk | 16.00 | 0.65 | 3.070 | −0.5000 |
| Oc | 16.00 | 0.65 | 3.070 | 0.0000 |
| Nam | 14.00 | 0.70 | 3.200 | 1.0000 |
| Ham | 1.00 | 0.00 | 0.000 | 0.0000 |
| Ow | 16.00 | 0.65 | 3.166 | −0.8476 |
| Hw | 1.00 | 0.00 | 0.000 | 0.4238 |

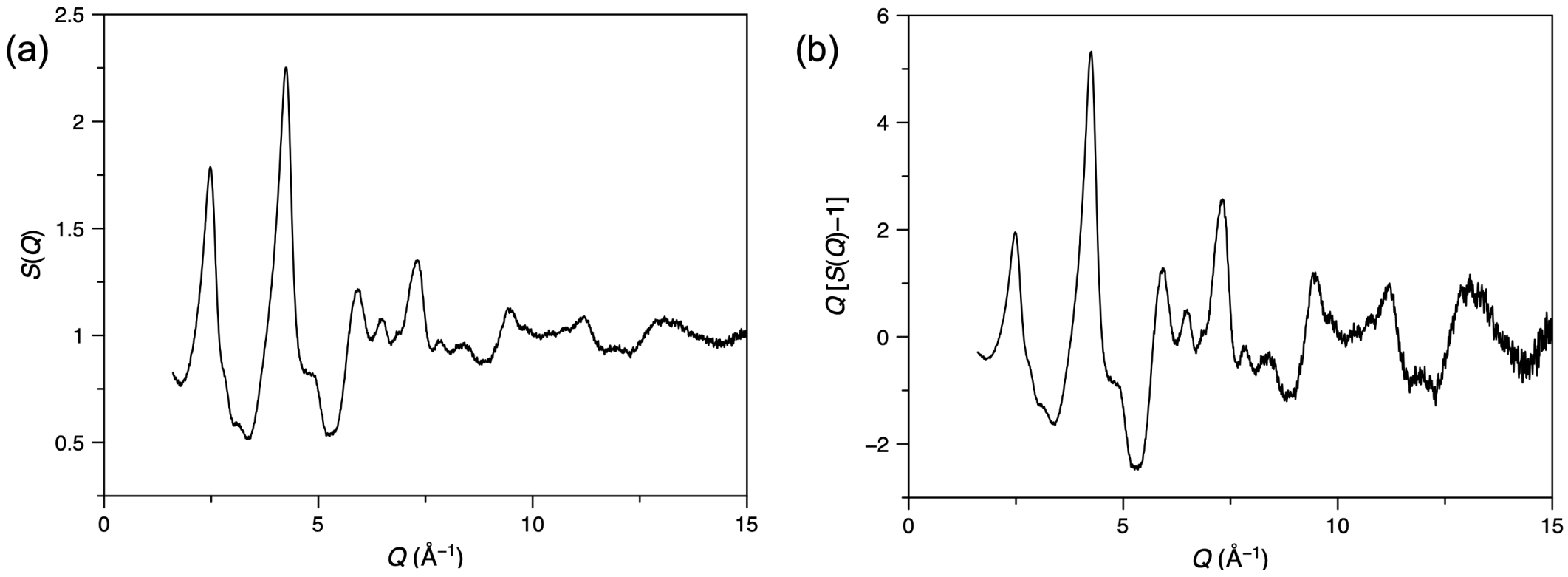

**Supplementary Figure 2** Average **(a)** $S(Q)$ and **(b)** $Q\ [S(Q)-1]$ obtained for ferrihydrite using PDFgetX3.

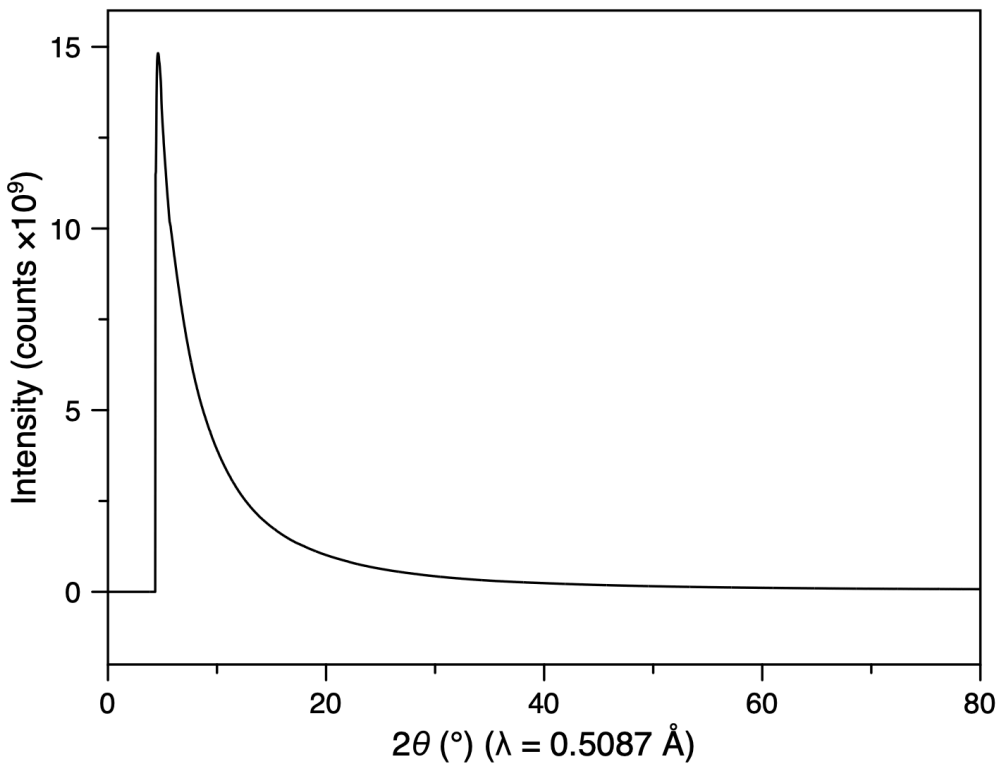

**Supplementary Figure 3** Averaged total scattering data obtained from the empty instrument (*i.e.* air scattering).

**Supplementary Table 4** Parameters obtained from refinement against the PDF for $CeO_2$ measured in a capillary.

| | **$CeO_2$** |
|---:|:---:|
| **$R_w$** | 0.27 |
| ***a*** **(Å)** | 5.4078(1) |
| **$B_{iso}$ Ce (Å$^2$)** | 0.33(2) |
| **$B_{iso}$ O (Å$^2$)** | 1.3(1) |
| **$Q_{damp}$ (Å$^{-1}$)** | 0.0094(7) |
| **$Q_{broad}$ (Å$^{-1}$)** | 0.011(1) |

**Supplementary Table 5** Parameters obtained from refinement against the PDF for Si measured in a capillary.

| | **Si** |
|---:|:---:|
| **$R_w$** | 0.28 |
| ***a*** **(Å)** | 5.4257(1) |
| **$B_{iso}$ Si (Å$^2$)** | 0.64(1) |

**Supplementary Table 6** Parameters obtained from refinement against the PDF for ferrihydrite measured in a capillary.

| | **Ferrihydrite** |
|---:|:---:|
| **$R_w$** | 0.39 |
| ***a*** **(Å)** | 5.95(1) |
| ***c*** **(Å)** | 9.01(3) |
| **$B_{iso}$ Fe (Å$^2$)** | 0.55(8) |
| **$B_{iso}$ O (Å$^2$)** | 1.3(3) |
| **$sp_{diameter}$ (Å)** | 16.4(7) |

**Supplementary Table 7** Parameters obtained from refinement against the PDF for Si measured in the liquid jet.

| | **Si in water** |
|---:|:---:|
| **$R_w$** | 0.38 |
| ***a*** **(Å)** | 5.44(2) |
| **$B_{iso}$ Si (Å$^2$)** | 1.4 (1.8) |

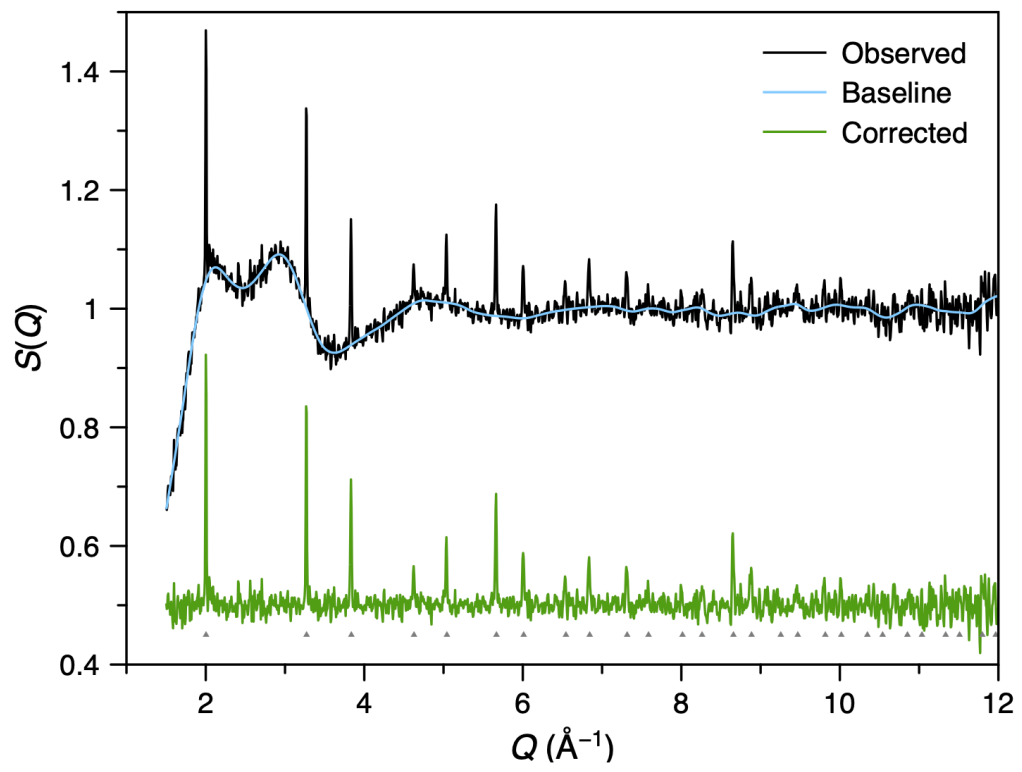

**Supplementary Figure 4** After subtracting the air scattering, the measured signal from Si in the liquid jet (black) contained the expected Si Bragg peaks (tick marks) superimposed on a residual signal from water scattering. A cubic smoothing spline was used to fit the baseline (blue) and remove the water signal. The resulting corrected data exhibit a flat baseline while retaining the Si Bragg peaks (green). The corrected data have been offset for clarity.

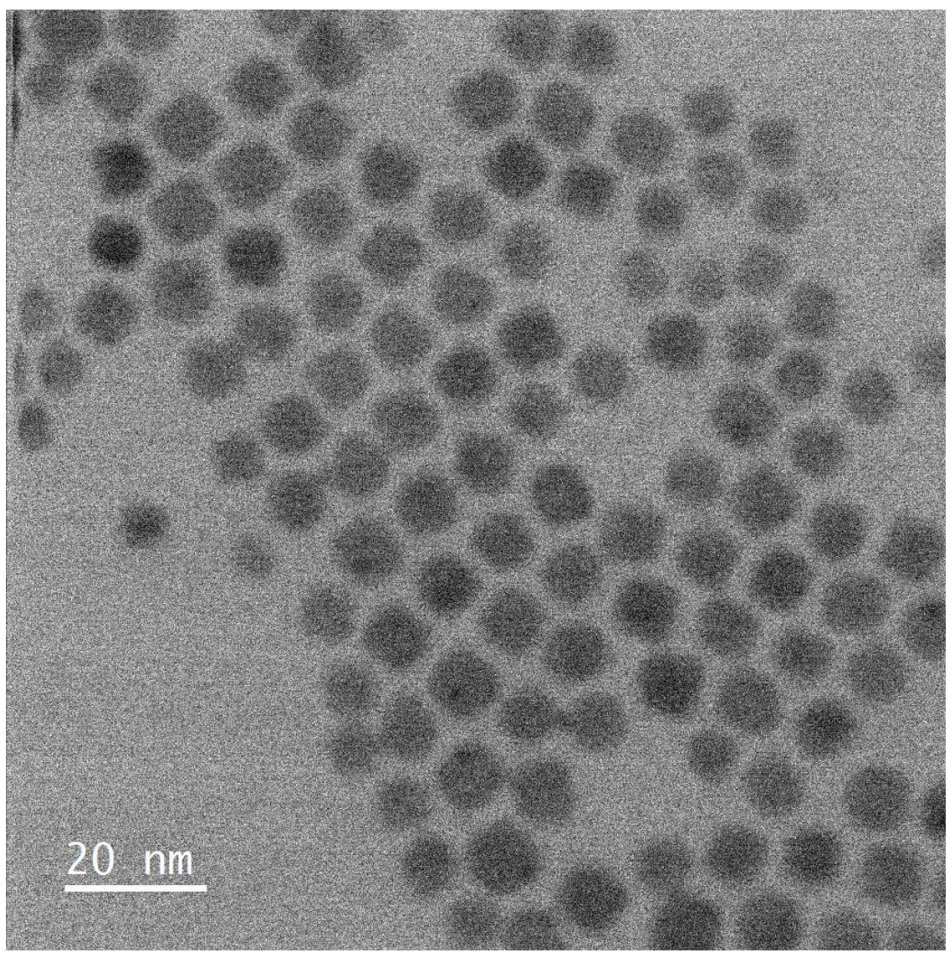

**Supplementary Figure 5** Transmission electron microscopy image of $CsPbBr_3$ nanocrystals.

**Supplementary Table 8** Equivalent cubic lattice parameter ($a_k = (abc/4)^{1/3}$ with $a$, $b$, and $c$ being the orthorhombic unit cell axes) and $B_{iso}$ refined for $CsPbBr_3$ colloidal nanocrystals at delay times of + and −5 ps under photoexcitation. For the datasets collected at −5 ps, the small variations in the refined isotropic strains were averaged, yielding an equivalent cubic unit cell parameter $a_0$ = 5.842(2) Å for the equilibrium structure, slightly expanded compared to that obtained from the ESRF data (5.838(2) Å). The Debye-Waller $B_{iso}$ factors of the equilibrium structure are Cs 4.6(4), Pb 1.6(1), $Br_{eq}$ 1.2(2), and $Br_{ax}$ 1.8(2) $Å^2$. The uncertainties of the refined parameters in the Table are of the same order of magnitude.

| Fluence ($mJ\ cm^{-2}$) | Delay time (ps) | $a_k$ (Å) | $B_{iso}$ (Cs) ($Å^2$) | $B_{iso}$ (Pb) ($Å^2$) | $B_{iso}$ ($Br_{eq}$) ($Å^2$) | $B_{iso}$ ($Br_{ax}$) ($Å^2$) |
|---|---|---|---|---|---|---|
| **0.82** | −5 | | 5.1 | 1.7 | 0.9 | 2.0 |
| | +5 | 5.841 | 5.2 | 1.6 | 1.2 | 1.8 |
| **1.97** | −5 | | 4.6 | 1.8 | 1.0 | 1.8 |
| | +5 | 5.845 | 5.2 | 2.1 | 1.2 | 1.8 |
| **3.94** | −5 | | 4.7 | 1.6 | 1.4 | 1.8 |
| | +5 | 5.850 | 5.7 | 2.1 | 1.3 | 2.1 |
| **5.77** | −5 | | 4.2 | 1.5 | 1.4 | 1.5 |
| | +5 | 5.851 | 4.9 | 1.8 | 1.4 | 1.2 |

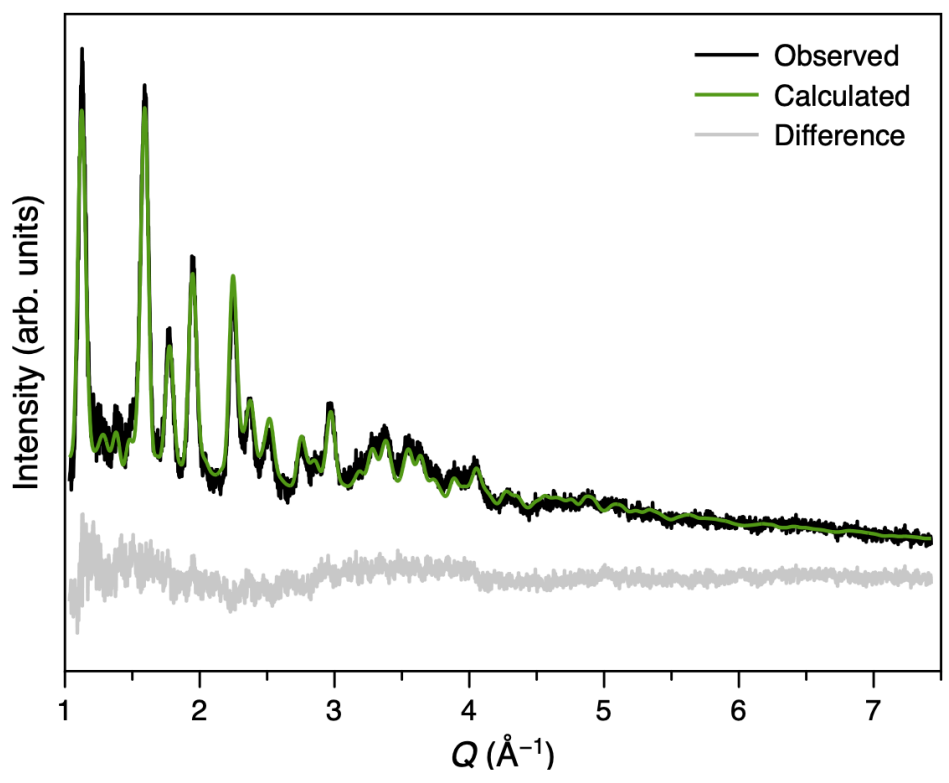

**Supplementary Figure 6** Debye scattering equation fit to $CsPbBr_3$ colloidal nanocrystals dispersed in toluene. The high-resolution synchrotron X-ray scattering data were collected at the ID22 beamline of the European Synchrotron Radiation Facility at 33 keV photon energy. The refined model was used as the structural reference for the analysis of the XFEL datasets, as detailed in the main text. For clarity, the experimental data are shown after subtraction of the solvent scattering contribution.

**Supplementary Table 9** Parameters obtained from refinement against the PDF for the Keggin cluster in aqueous solution measured in the liquid jet.

| | **Keggin** |
|---|---|
| $R_w$ | 0.20 |
| **Zoom scale** | 1.007(8) |
| $B_{iso}$ **W** ($Å^2$) | 0.23(4) |
| $B_{iso}$ **O** ($Å^2$) | – |